\documentclass[aps,prl,twocolumn,superscriptaddress]{revtex4-1}

\usepackage{graphicx}
\usepackage{units}

\usepackage{color}

\usepackage{caption}
\DeclareCaptionLabelSeparator{dot}{. }
\makeatletter
\def\justified{
	\let\\\@normalcr
	\@rightskip\z@skip \rightskip\@rightskip
	\leftskip\z@skip
	\parindent 0em\relax
	\setlength{\parfillskip}{0pt plus 1fil}}
\DeclareCaptionJustification{justified}{\justified}
\usepackage{subcaption}
\usepackage{times}
\usepackage{multirow}
\usepackage{amssymb}
\usepackage{dcolumn}
\usepackage{bm}
\usepackage{amsmath}
\usepackage{fancyhdr}
\usepackage{float}
\usepackage{esvect}
\usepackage{setspace}
\usepackage{braket}
\usepackage{here}
\usepackage{hyperref}
\hypersetup{colorlinks=true, linkcolor=blue,citecolor=blue,urlcolor=blue}
\def\OMIT#1 {{}}
\def\MEMO#1 {{}}

\newcommand{\Er}{{^{167}}{\rm Er}}
\newcommand{\mnineteen}{|\text{--}19/2\rangle}
\newcommand{\mseventeen}{|\text{--}17/2\rangle}
\newcommand{\mfifteen}{|\text{--}15/2\rangle}
\newcommand{\mthirteen}{|\text{--}13/2\rangle}

\newcommand{\mnine}{|\text{--}9/2\rangle}
\newcommand{\mseven}{|\text{--}7/2\rangle}
\newcommand{\mfive}{|\text{--}5/2\rangle}

\newcommand{\mone}{|\text{--}1/2\rangle}

\newcommand{\pthirteen}{|13/2\rangle}

\newcommand{\pnine}{|9/2\rangle}

\newcommand{\pfive}{|5/2\rangle}
\newcommand{\pthree}{|3/2\rangle}
\newcommand{\pone}{|1/2\rangle}
\newcommand{\mfstate}{|m_F \rangle}

\begin{document} 
\title{Controlling Dipolar Exchange Interactions in a Dense 3D Array of Large Spin Fermions}

\author
{A.\,Patscheider}
\affiliation{
		Institut f\"ur Experimentalphysik, Universit\"at Innsbruck, Technikerstra{\ss}e 25, 6020 Innsbruck, Austria}
		
\author
{B.\,Zhu}
\affiliation{
		JILA, NIST, and Department of Physics,
        University of Colorado, 440 UCB, Boulder, CO 80309, USA}
\affiliation{
		ITAMP, Harvard-Smithsonian Center for Astrophysics, Cambridge, MA 02138, USA}
\affiliation{
		Department of Physics, Harvard University, Cambridge, MA 02138, USA}

\author
{L.\,Chomaz}
\affiliation{
		Institut f\"ur Experimentalphysik, Universit\"at Innsbruck, Technikerstra{\ss}e 25, 6020 Innsbruck, Austria}		

\author
{D.\,Petter}
\affiliation{
		Institut f\"ur Experimentalphysik, Universit\"at Innsbruck, Technikerstra{\ss}e 25, 6020 Innsbruck, Austria}

\author
{S.\,Baier}
\affiliation{
		Institut f\"ur Experimentalphysik, Universit\"at Innsbruck, Technikerstra{\ss}e 25, 6020 Innsbruck, Austria}

\author
{A-M.\,Rey}
\affiliation{
		JILA, NIST, and Department of Physics,
        University of Colorado, 440 UCB, Boulder, CO 80309, USA}

\author
{F.\,Ferlaino}
\affiliation{
		Institut f\"ur Experimentalphysik, Universit\"at Innsbruck, Technikerstra{\ss}e 25, 6020 Innsbruck, Austria}
\affiliation{
		Institut f\"ur Quantenoptik und Quanteninformation, \"Osterreichische Akademie der Wissenschaften, Technikerstra{\ss}e 21a, 6020 Innsbruck, Austria}

\author
{M.\,J.\,Mark}
\affiliation{
		Institut f\"ur Experimentalphysik, Universit\"at Innsbruck, Technikerstra{\ss}e 25, 6020 Innsbruck, Austria}
\affiliation{
		Institut f\"ur Quantenoptik und Quanteninformation, \"Osterreichische Akademie der Wissenschaften, Technikerstra{\ss}e 21a, 6020 Innsbruck, Austria}

\date{April 2019}

\begin{abstract}
Dipolar interactions are ubiquitous in nature and rule the behavior of a broad range of systems spanning from energy transfer in biological systems to quantum magnetism. Here, we study magnetization-conserving dipolar induced spin-exchange dynamics in dense arrays of fermionic erbium atoms confined in a deep three-dimensional lattice. Harnessing the special atomic properties of erbium, we demonstrate control over the spin dynamics by tuning the dipole orientation and changing the initial spin state within the large 20 spin hyperfine manifold. Furthermore, we demonstrate the capability to quickly turn on and off the dipolar exchange dynamics via optical control. The experimental observations are in excellent quantitative agreement with numerical calculations based on discrete phase-space methods, which capture entanglement and beyond-mean field effects. Our experiment sets the stage for future explorations of rich magnetic behaviors in long-range interacting dipoles, including exotic phases of matter and applications for quantum information processing.
\end{abstract}

\maketitle

Spin lattice models of localized magnetic moments (spins), which interact with one another via exchange interactions, are paradigmatic examples of strongly correlated many-body quantum systems. Their implementation in clean, isolated, and fully controllable lattice confined ultra-cold atoms opens a path for a new generation of synthetic quantum magnets, featuring highly entangled states, especially when driven out-of-equilibrium, with broad applications ranging from precision sensing and navigation, to quantum simulation and quantum information processing~\cite{Bloch2008,Gross2017}. However, the extremely small energy scales associated with the nearest-neighbor spin interactions in lattice-confined atoms with dominant contact interactions, have made the observation of quantum magnetic behaviors extremely challenging~\cite{Bloch2008r,Greif2013}. On the contrary, even under frozen motional conditions, dipolar gases, featuring long-range and anisotropic interactions, offer the opportunity to bring ultra-cold systems several steps ahead towards the ambitious attempt of modeling and understanding quantum magnetism. While great progress in studying quantum magnetism has been achieved using arrays of Rydberg atoms \cite{Zeiher2017,Bernien2017,Barredo2018,Guardado2018}, trapped ions \cite{Neyenhuise2017,Blatt2012,Britton2012}, polar molecules \cite{Yan2013ood,Hazzard2014}, and spin-$3$ bosonic chromium atoms~\cite{dePaz2013,dePaz2016,Lepoutre2018}, most of the studies so far have been limited to spin-$1/2$ mesoscopic arrays of at the most few hundred particles or to macroscopic but dilute ($<0.1$ filling fractions) samples of molecules in lattices.

\begin{figure*}[t!]
	\includegraphics[width=0.95\textwidth]{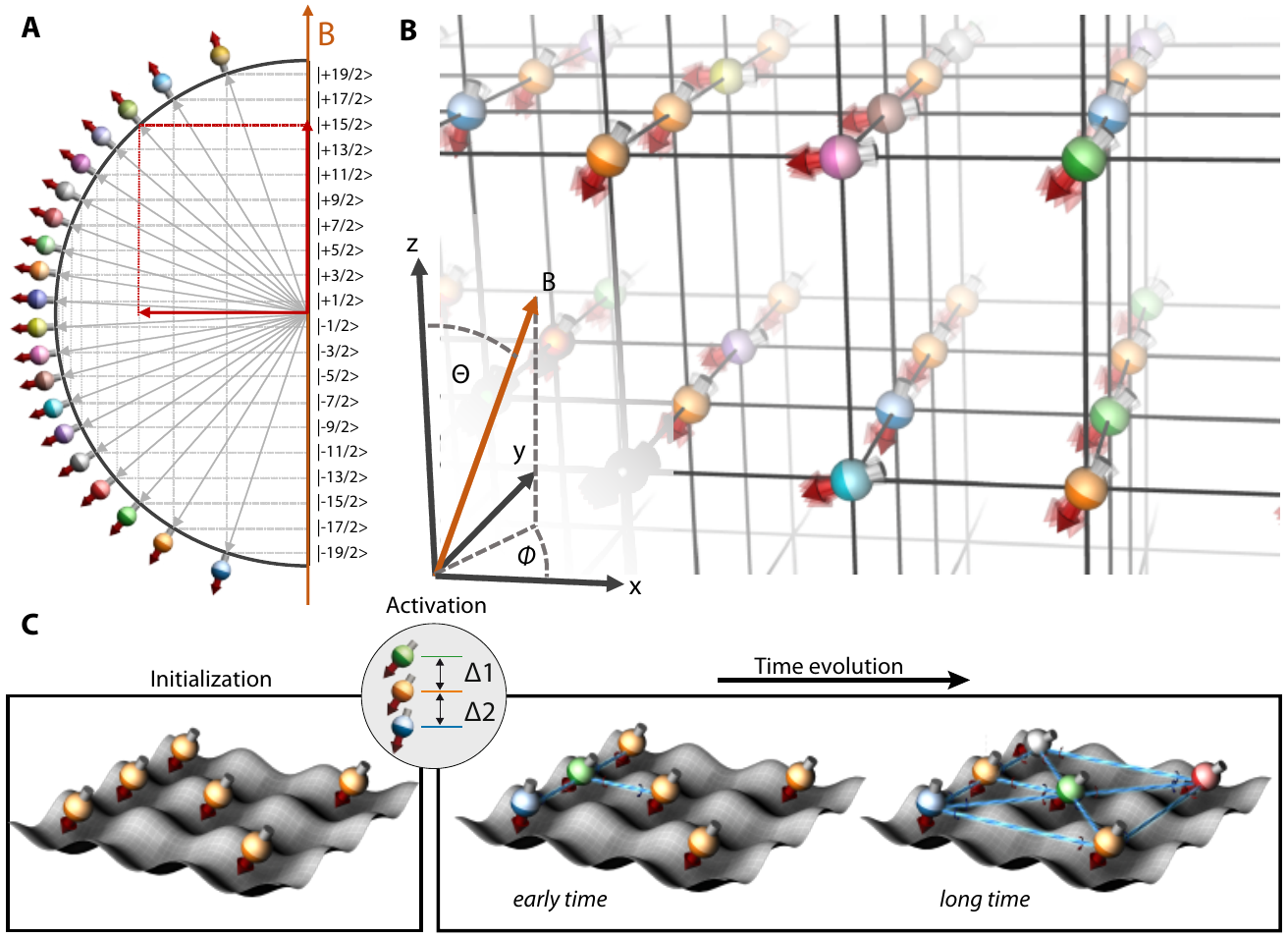}
\caption{Overview of the experimental setup. (A) Illustration of the total spin space of a single $\Er$ atom in the lowest hyperfine level $|F=19/2\rangle$ with all $20\,m_F$ states. The angle of the symbols indicates the orientation of the total spin $|\mathbf{F}|=\sqrt{F(F+1)}$ in relation to the quantization axis. (B) Sketch of the experimental system, an anisotropic 3D lattice structure filled with fermionic $^{167}$Er with a quantization axis tunable by the angles ($\Theta,\phi$) of the external magnetic field $B$. (C) Illustration of the experimental sequence (from left to right): The system is initialized by preparing all atoms in one starting state, here $\mseventeen$. We activate the spin dynamics by changing the magnetic field to set $\bar{\delta}=0$. On early time scale dynamics are happening mainly among nearest neighbor atoms. Subsequently interactions between atoms at larger distances are involved in the dynamics.}
\label{fig:1}
\end{figure*}

In this work, we report the first investigations of non-equilibrium quantum magnetism in a dense array of fermionic magnetic atoms confined in a deep three-dimensional optical lattice. Our platform realizes a quantum simulator of the long-range XXZ Heisenberg model. The simulator roots on the special atomic properties of ${}^{167}$Er, whose ground state bears large angular momentum quantum numbers with $I=7/2$ for the nuclear spin and $J=6$ for the electronic angular momentum, resulting in a $F=19/2$ hyperfine manifold, as depicted in Fig.\,\ref{fig:1}A. Such a complexity enables new control knobs for quantum simulations. First, it is responsible for the large magnetic moment in Er. Second, it gives access to a fully controllable landscape of $20$ internal levels, all coupled by strong magnetic dipolar interactions, up to $49$ times larger than the ones felt by $F=1/2$ alkali atoms in the same lattice potential~\cite{Stamper-Kurn2013}. Finally, it allows fast optical control of the energy splitting between the internal states which can be tuned on and off resonance using the large tensorial light shift~\cite{Becher2017pol}, which adds to the usual quadratic Zeeman shift.

Using all these control knobs, we explore the dipolar exchange dynamics and benchmark our simulator with an advanced theoretical model, which takes entanglement and beyond mean-field effects into account~\cite{suppmat}. In particular, we initialize the system into a desired spin state and activate the spin dynamics, while the motional degree of freedom mainly remains frozen. Here, we study the spreading of the spin population in the different magnetic sublevels as a function of both the specific initial spin state and the dipole orientation. We demonstrate that the spin dynamics at short evolution time follows a scaling that is invariant under internal state initialization (choice of macroscopically populated initial Zeeman level) and that is set by the effective strength of the dipolar coupling. On the contrary, at longer times, the many-body dynamics is affected by the accessible spin space and the long-range character of dipolar interactions beyond nearest neighbors. Finally, we show temporal control of the exchange dynamics using off resonant laser light.

The XXZ Heisenberg model that rules the magnetization-conserving spin dynamics of our system can be conveniently written  using spin-$19/2$ dimensionless angular momentum operators  $\hat{\mathbf{F}}_i=\{\hat{F}^x_i,\hat{F}^y_i,\hat{F}^z_i\}$, acting on site $i$ and satisfying the commutation relation $[\hat F_i^x,\hat F_i^y]=i\hat F_i^z$. We use the eigenbasis of $\hat{F}^z$ denoted as $|m_F\rangle$ with $0\leq |m_F|\leq F$~\,\cite{Auerbach1994iea,Dutta2015,suppmat}: 

\begin{eqnarray} 
\hat{H} &=& \frac{1}{2}\sum_{i,j\neq i}V_{i,j} \left(\hat{F}_ {i}^z\hat{F}_{j}^z-\frac{1}{4}(\hat{F}^+_i\hat{F}^-_{j}+\hat{F}^-_{i}\hat{F}^+_{j}) \right) \nonumber \\ &&+ \sum_{i} \delta_{i}(\hat{F}^z_{i} )^2
\end{eqnarray} 

The coupling constants $V_{i, j}\,=\,V_{dd}d_y^3\frac{1-3\cos^2(\theta_{i,j})}{r_{ij}^3}$, describe the direct dipole-dipole interactions (DDI), which have long-range character and thus couple beyond nearest neighbors. The dipolar coupling strength between two dipoles located at $\vec{r}_{i}$ and $\vec{r}_{j}$ depends on their relative distance $r_{ij}=|\vec{r}_{i}-\vec{r}_{j}|$ and on their orientation, described by the angle $\theta_{i,j}$ between the dipolar axis, set by the external magnetic field, and the interparticle axis; see Fig.\,\ref{fig:1}B.  Here, $V_{dd}\,\approx\,\frac{\mu_0g_F^2\mu_B^2}{4\pi d_y^3}$ denotes the dipolar coupling strength, with $g_F\approx 0.735$ for ${}^{167}$Er, $\mu_0$  the magnetic permeability of vacuum,  $\mu_B$ the Bohr magneton, and $d_y$ the shortest lattice constant. The $\hat{F}_ {i}^z\hat{F}_{j}^z$ terms in the Hamiltonian account for the diagonal part of the interactions while the $\hat{F}^+_i\hat{F}^-_{j}+\hat{F}^-_{i}\hat{F}^+_{j}$ terms describe dipolar exchange processes. The second sum denotes the single particle quadratic term $\delta_{i}(\hat{F}^z_{i})^2$ with $\delta_{i}=\delta_{i}^Z+\delta_{i}^T$, accounting for the quadratic Zeeman effect $\propto \delta_{i}^Z$ and tensorial light shifts  $\propto \delta_{i}^T$. These two contributions can be independently controlled in our experiment.

The quadratic Zeeman shift allows us to selectively prepare all atoms in one target state of the spin manifold~\cite{suppmat}. The tensorial light shift can compete or cooperate with the quadratic Zeeman shift and can be used as an additional control knob to activate/deactivate the exchange processes. Note that, for all measurements, a large linear Zeeman shift is always present, but since it does not influence the spin-conserving dynamics, it is omitted in Eq.\,1.

In the experiment, we first load a spin-polarized quantum degenerate Fermi gas of $\approx 10^4$ Er atoms into a deep 3D optical lattice, following the scheme of Ref.~\cite{Baier2017sif}. The cuboid lattice geometry with lattice constants $(d_x,d_y,d_z) = (272,266,544)\,$nm results in weakly coupled 2D planes, with typical tunneling rates of $\sim10\,$Hz inside the planes and $\sim\,$mHz between them~\cite{suppmat}. The external magnetic field orientation, setting the quantization axis as well as the dipolar coupling strengths, is defined by the polar angles $\Theta$ and $\phi$ in the laboratory frame; see Fig.\,\ref{fig:1}B. The fermionic statistics of the atoms enables us to prepare a dense band insulator with at most one atom per lattice site, as required for a clean implementation of the XXZ Heisenberg model. This is an advantage of fermionic atoms as compared to bosonic systems, which typically require filtering protocols to remove doublons~\cite{Lepoutre2018}. 

\begin{figure*}[t!]
\includegraphics[width=0.95\linewidth]{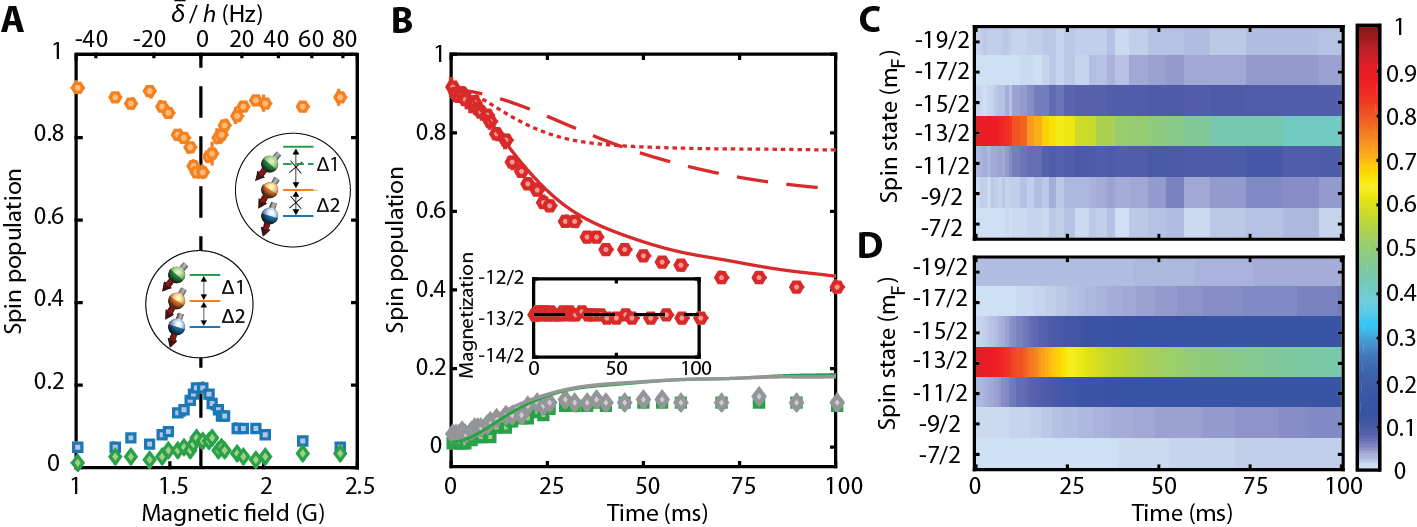}
\caption{ Spin exchange dynamics. (A) Measured spin populations in states $m_F=\mseventeen$ (circles), $m_F\pm1$ ( diamonds and squares) after $50\,$ms hold time as a function of the magnetic field with $\Theta=0^\circ$. A Gaussian fit (not shown) to the data provides a resonant magnetic field value of $\approx 1.67\,$G. The top axis shows the mean total detuning $\bar{\delta}$ from the resonance condition. (B) Measured spin population in states $m_F=\mthirteen$ (circles), $m_F\pm1$ (diamonds and squares) as a function of the hold time after quenching onto the spin exchange resonance with $\Theta=0^\circ$. The dashed line shows the simple mean-field result, the dotted line gives the NNI-GDTWA result, and the solid lines denotes the full-GDTWA result. (C-D) Spin diffusion with initial state $|m_F^0 \rangle=\mthirteen$ plotting the population of states from $m_F-3$ to $m_F+3$ as a function of the hold time, for the experiment (C) and the full-GDTWA model (D), with the same initial conditions as (B). Datapoints consist of a minimum of 4 individual realizations, error bars denote the standard error of the mean.}
\label{fig:2}
\end{figure*}

Our experimental sequence to study the spin dynamics is illustrated in Fig.\,\ref{fig:1}C. In particular, we prepare the system into the targeted $m_F^0$ state by using the lattice-protection protocol demonstrated in Ref.\,\cite{Baier2017sif}; see also Ref.\,\cite{suppmat}. At the end of the preparation, the majority of atoms are in the desired $m_F^0$ ($>80\%$) at $B\approx 4\,$G. We note that atom losses during the spin preparation stage reduces the filling factor to about 60\% of the initial one~\cite{suppmat}. We then activate the spin dynamics by quenching the magnetic field to a value for which $\bar{\delta}=\sum_{i} \delta_{i} =0$, providing a resonance condition for the magnetization-conserving spin-exchange processes; see Fig.\,\ref{fig:2}A. After a desired time of evolution, we stop the dynamics by rapidly increasing the magnetic field, leaving the resonance condition. We finally extract the atom number in each spin state via a spin-resolved band-mapping technique~\cite{suppmat} and derive the relative state populations by normalization to the initial total atom number.

We now probe the evolution of the spin-state population as a function of the hold time on resonance. We observe a redistribution of the population from the initial state to multiple neighboring states in $m_F$ space, as exemplary shown for an initial state of $\mthirteen$ in Fig.\,\ref{fig:2}B-C. The dynamics preserves the total magnetization; see inset of Fig.\,\ref{fig:2}B. We observe similar behavior independently of the initialized $m_F^0$ states. The spin transfer happens sequentially. At short times it is dominated by the transfer to states directly coupled by the dipolar exchange Hamiltonian, i.\,e.\,those ones which differ by plus/minus one unit of angular momentum ($\Delta m_F = \pm 1$). At longer times, subsequent processes transfer atoms to states with $|\Delta m_F| \geq 2$; see Fig.\,\ref{fig:2}C-D. 

To benchmark our quantum simulator, we use a semiclassical phase-space sampling method, the so called generalized discrete truncated Wigner approximation (GDTWA)~\cite{Lepoutre2018,Schachenmayer2015a,Schachenmayer2015b,Polkovnikov2010,suppmat}. The method accounts for quantum correlation in the many-body dynamics and is  adapted to tackle the complex dynamics of a large-spin system in a regime where exact diagonalization techniques are impossible to implement with current computers. The GDTWA calculations take into account actual experimental parameters such as spatial inhomogeneites, density distribution after the lattice loading, initial spin distribution, and effective lattice filling, including the loss during the spin preparation protocol~\cite{suppmat}. Figure~\ref{fig:2}B shows the experimental dynamics together with the GDTWA simulations. Although the model does not include corrections due to losses and tunneling during the dynamics, it successfully captures the behavior of our dense system not only at short time, but also at long time, where the population dynamics slows down and starts to reach an equilibrium. Similar level of agreement between experiment and theory is shown in Fig.\,\ref{fig:2}C-D where we directly compare the spreading of the spin population as a function of time. 

The important role of quantum effects in the observed spin dynamics can be illustrated by contrasting the GDTWA simulation with a mean-field calculation. Indeed, the mean-field calculation fails in capturing the system behavior. It predicts a too slow population dynamics for non-perfect spin-state initialization, as in the experiments shown in Fig.\,\ref{fig:2}B, and no dynamics for the ideal case where all atoms are prepared in the same internal state~\cite{suppmat}. To emphasize the beyond nearest-neighbor effects, we also compare the experiment with a numerical simulation that only includes nearest-neighbor interactions (NNI-GDTWA). Here, we again observe a very slow spin evolution, which largely deviates from the measurements. The agreement of the full GDTWA predictions with our experimental observations points to the long-range many-body nature of the underlying time evolution. Our theory calculations also support the built up of entanglement during the observed time evolution.

\begin{figure*}[t!]
\includegraphics[width=0.95\linewidth]{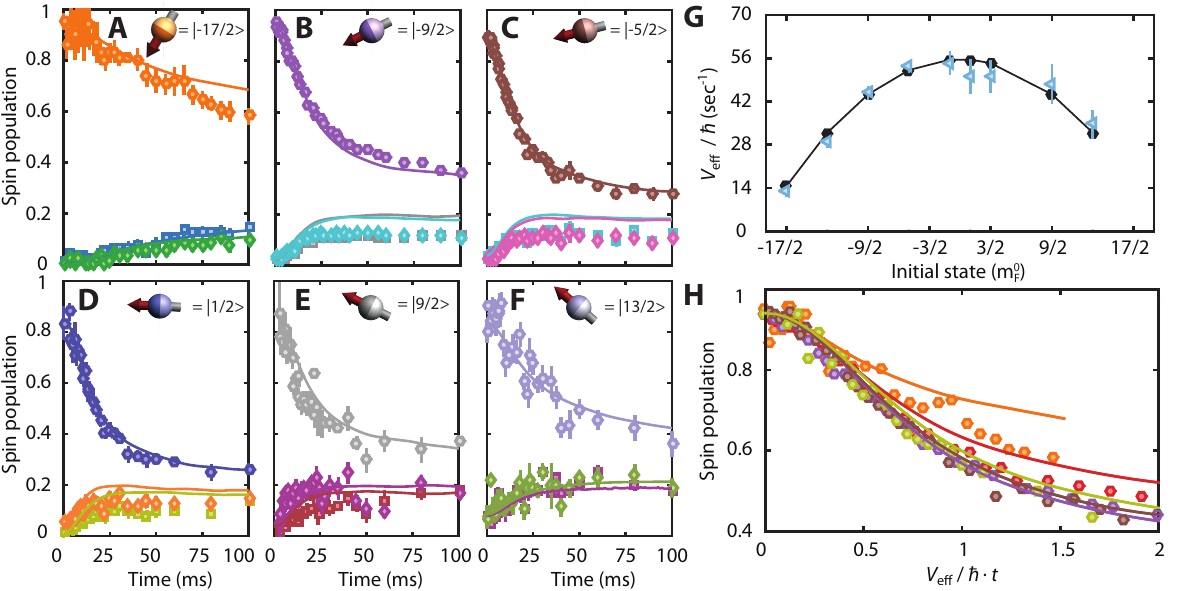}
\caption{State dependence of the spin exchange dynamics. (A-F) Dynamic evolution of the initial states $\mseventeen$ (A), $\mnine$ (B), $\mfive$ (C), $\pone$ (D), $\pnine$ (E), and $\pthirteen$ (F) and of the corresponding neighboring states $m_F\pm1$ together with the full-GDTWA results (solid lines) for $\Theta =0^{\circ}$. (G) Extracted $V_\text{eff}$ as a function of the initial state $m_F^0$ from a fit to the experimental data (cyan triangles) and numerically computed from the initial spin distribution (black circles). Errorbars denote the $68\%$ confidence interval of the fits. (H) All datasets with $m_F^0<0$ used in (G) together with the corresponding full-GDTWA results (solid lines) plotted in units of the rescaled time $\tau = V_\text{eff}/\hbar\cdot t$. Note that all experiment and theory data are plotted for times, $t \leq 100\,$ms, of (A-F). To account for the different preparation fidelity, the populations of the initial states are shifted to $1$ by adding a constant offset. For clarity error bars are omitted here.}
\label{fig:3}
\end{figure*}

Different spin configurations feature distinct effective interaction strengths, which also depend on the orientation of the dipoles with respect to the lattice. We demonstrate our ability to control this interaction, which governs the rate of population exchange, by the choice of the initial $m_F^0$ state and the orientation of the external magnetic field. This capability provides us with two tuning knobs to manipulate dipolar exchange interactions in our quantum simulator. Figure~\ref{fig:3}A-F plots the dynamics of the populations for three neighboring spin states after the quench, starting from different initial spin states. Solid lines show the results of the full-GDTWA calculations. For each initial $m_F^0$, we find a remarkable agreement between theory and experiment. We observe a strong speedup for states with large spin projections perpendicular to the quantization axis, as it is expected from the expectation value of $\hat{F}^+_i\hat{F}^-_j$, which gives a prefactor  $\gamma(m_F^0)=\sqrt{F(F+1)-m_F^0(m_F^0+1)}\sqrt{F(F+1)-m_F^0(m_F^0-1)}$. The initial dynamics can be well described by a perturbative expansion up to the second order~\cite{suppmat}, resulting in the analytic expression for the normalized population $n_{m_F}(t)$ of the initial state:
\begin{eqnarray}
  n_{m_F^0}(t)&=&n_{m_F^0}(0)\Big(1-n_{m_F^0}(0)\frac{V_{\rm eff}^2}{\hbar^2}t^2\Big)
\end{eqnarray}

Here, $V^2_\text{eff}\equiv \frac{\gamma^2(m_F^0)}{8 N}\sum_{i,j\neq i} V_{ij}^2$ is the overall effective interaction strength summed over $N$ atoms and $n_{m_F^0}(0)$ denotes the purity of the initial state preparation. For a quantitative analysis of the early-time spin evolution, we compare the theoretically calculated $V_\text{eff}$ from the initial atomic distribution used in the GDTWA model with the one extracted from a fit of Eq.\,2 to the experimental data.
Here we consider the data up to $t<0.5\frac{\hbar}{V_\text{eff}}$ estimated using the theoretically calculated $V_\text{eff}$~\cite{Note1}. Figure~\ref{fig:3}G plots both, the theoretical and experimental $V_\text{eff}$ as a function of the initial $m_F^0$ and highlights once more their quantitative agreement. The interaction parameter $V_\text{eff}$, can also be used to rescale the time axis. As shown in Fig.\,\ref{fig:3}H, all data sets now collapse onto each other for $\frac{t V_\text{eff}}{\hbar}<0.5$, revealing the invariant character of the short-time dynamics under the internal state initialization. At longer timescales, the theory shows that the timetraces start to deviate from each others and saturate to different values, indicating that thermal-like states are on reach. In the experiment, we observe a similar behavior but here the saturation value might also be affected by losses and residual tunneling.

\begin{figure}[t!]
\includegraphics[width=0.95\linewidth]{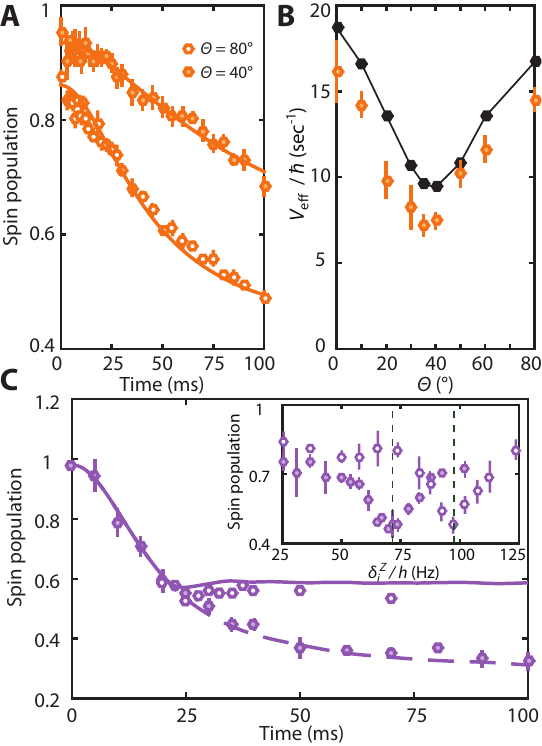}
\caption{Angle dependence of the spin exchange dynamics and dynamical control. (A) Exemplary measurements of the time evolution for the starting spin state $\mseventeen$ for $\Theta\,=\,40^\circ,80^\circ$. Solid lines show the full-GDTWA results. (B) Extracted $V_\text{eff}$ as a function of $\Theta$ from a fit to the experimental data (orange circles) and numerically computed from the initial spin distribution (black circles). Errorbars denote the $68\%$ confidence interval of the fits. (C) Time evolution of the initial state $\mnine$ at $\bar \delta = 0$ and $\Theta = 0^\circ$ without (filled circles) and with (open circles) switching on an additional light field after $20\, \rm ms$ of evolution. Solid (dashed) lines are the corresponding full-GDTWA calculations. The inset shows the population of the initial spin state after $50\, \rm ms$ evolution time as a function of the quadratic Zeeman shift without (filled circles) and with (open circles) the additional light field. Determining the centers of the resonances via a fit yields an absolute shift of the resonance condition by $h \times 27(1)\, \rm Hz$ between both conditions.}
\label{fig:4}
\end{figure}

Because of the anisotropic character of the DDI, the strength of the dipolar exchange can be controlled by changing the angle $\Theta$; see Fig.\,\ref{fig:1}B. As exemplary shown in Fig.\,\ref{fig:4}A for $\mseventeen$, the observed evolution speed of the spin populations strongly depends on $\Theta$, changing by about a factor of $2$ between $\Theta=40^\circ$ and $80^\circ$. The GDTWA results show a very good quantitative agreement with the experiment. 
We repeat the above measurements for different values of $\Theta$ and we extract $V_\text{eff}$; Fig.\,\ref{fig:4}B. 
It is worth to notice that, while the dipolar interactions can be completely switched off at a given angle in a 1D chain, in a 3D system the situation is more complicated. However, as expected by geometrical arguments, we observe that the overall exchange strength becomes minimal for a specific dipole orientation $(\Theta_c\approx 35^\circ,\phi_c=45^\circ)$.
We compare our measured $V_\text{eff}$ with the ones calculated from the initial spin distribution, which is a good quantity to describe the early time dynamics. 
Theory and experiment show a similar trend, in particular reaching a minimum at about $\Theta_c$. 
Note that the simple analytic formula (Eq.\,2), used for fitting the data, deviates from the actual evolution at longer times. This leads to a small down-shift of the experimental values~\cite{suppmat}.

Finally, we demonstrate fast optical control of the spin dynamics relying on the remarkably large tensorial light shift of erbium compared to alkali atoms. As shown in Fig.\,\ref{fig:4}C, we can almost fully suppress the spin exchange dynamics by suddenly switching on a homogeneous light field after an initial evolution time on resonance. Therefore the tensorial light shift, inducing a detuning from the resonance condition (see inset), allows a full spatial and temporal control over the exchange processes as the light power can be changed orders of magnitude faster than the typical interaction times and can address even single lattice sites in quantum gas microscopes. This capability can be an  excellent resource for quantum information processing, i.\,e.\,we could use dipolar exchange processes to efficiently prepare highly entangled states between different parts of a quantum system and then store the quantum information at longer times by turning the interactions off. 

The excellent agreement between the experiment and the theory, not only verifies our quantum simulator but sets the stage towards its use for the study of new regimes intractable to theory. For example by operating at weaker lattice when motion is involved, the dynamics is no longer described by a spin model but by a high spin Fermi-Hubbard model with long-range interactions. Alternatively by treating the internal hyperfine levels as a synthetic dimension \cite{Mancini1510} while adding Raman transitions to couple them, one could engineer non-trivial synthetic gauge field models even when tunneling is only allowed in one direction. Moreover, the demonstrated control over the different hyperfine level structure, our capability to tune the strength of the direct dipolar exchange coupling via the magnetic field angle, and the possibility of the dynamical and spatial control of the resonance condition via tensorial light shifts make our system a potential resource for quantum information processing with a qudit-type multi-level encoding using the 20 different interconnected hyperfine levels~\cite{Lanyon2008,Brion2007,Kues2017}. 

\begin{acknowledgments}
We thank J. Schachenmayer  for fruitful discussions and Arghavan Safavi-Naini for helping us understanding the loading into the lattice. We thank J.\,H.\,Becher and G.\,Natale for their help in the experimental measurements and for fruitful discussions. We also thank Rahul  Nandkishore  and Itamar Kimchi for reviewing the manuscript. The Innsbruck group is supported through an ERC Consolidator Grant (RARE, no.\,681432) and a FET Proactive project (RySQ, no.\,640378) of the EU H2020, and by a Forschergruppe (FOR 2247/PI2790) of the DFG and the FWF. LC is supported within a Marie Curie Project (DipPhase, no.\,706809) of the EU H2020. A.M.R is supported by the AFOSR grant FA9550-18-1-0319 and its MURI Initiative, by the DARPA and ARO grant W911NF-16-1-0576, DARPA-DRINQs, the ARO single investigator award W911NF-19-1-0210, the NSF PHY1820885, JILA-NSF PFC-173400 grants, and by NIST. B.Z. is supported by the NSF through a grant to ITAMP.
\end{acknowledgments}

* Correspondence and requests for materials
should be addressed to M.\,J.\,M.~(email: manfred.mark@uibk.ac.at).

\bibliography{Spindynamics}

\begin{thebibliography}{33}%
\makeatletter
\providecommand \@ifxundefined [1]{%
 \@ifx{#1\undefined}
}%
\providecommand \@ifnum [1]{%
 \ifnum #1\expandafter \@firstoftwo
 \else \expandafter \@secondoftwo
 \fi
}%
\providecommand \@ifx [1]{%
 \ifx #1\expandafter \@firstoftwo
 \else \expandafter \@secondoftwo
 \fi
}%
\providecommand \natexlab [1]{#1}%
\providecommand \enquote  [1]{``#1''}%
\providecommand \bibnamefont  [1]{#1}%
\providecommand \bibfnamefont [1]{#1}%
\providecommand \citenamefont [1]{#1}%
\providecommand \href@noop [0]{\@secondoftwo}%
\providecommand \href [0]{\begingroup \@sanitize@url \@href}%
\providecommand \@href[1]{\@@startlink{#1}\@@href}%
\providecommand \@@href[1]{\endgroup#1\@@endlink}%
\providecommand \@sanitize@url [0]{\catcode `\\12\catcode `\$12\catcode
  `\&12\catcode `\#12\catcode `\^12\catcode `\_12\catcode `\%12\relax}%
\providecommand \@@startlink[1]{}%
\providecommand \@@endlink[0]{}%
\providecommand \url  [0]{\begingroup\@sanitize@url \@url }%
\providecommand \@url [1]{\endgroup\@href {#1}{\urlprefix }}%
\providecommand \urlprefix  [0]{URL }%
\providecommand \Eprint [0]{\href }%
\providecommand \doibase [0]{http://dx.doi.org/}%
\providecommand \selectlanguage [0]{\@gobble}%
\providecommand \bibinfo  [0]{\@secondoftwo}%
\providecommand \bibfield  [0]{\@secondoftwo}%
\providecommand \translation [1]{[#1]}%
\providecommand \BibitemOpen [0]{}%
\providecommand \bibitemStop [0]{}%
\providecommand \bibitemNoStop [0]{.\EOS\space}%
\providecommand \EOS [0]{\spacefactor3000\relax}%
\providecommand \BibitemShut  [1]{\csname bibitem#1\endcsname}%
\let\auto@bib@innerbib\@empty
%</preamble>
\bibitem [{\citenamefont {Bloch}(2008)}]{Bloch2008}%
  \BibitemOpen
  \bibfield  {author} {\bibinfo {author} {\bibfnamefont {I.}~\bibnamefont
  {Bloch}},\ }\href {https://doi.org/10.1038/nature07126} {\bibfield  {journal}
  {\bibinfo  {journal} {Nature}\ }\textbf {\bibinfo {volume} {453}},\ \bibinfo
  {pages} {1016} (\bibinfo {year} {2008})}\BibitemShut {NoStop}%
\bibitem [{\citenamefont {Gross}\ and\ \citenamefont
  {Bloch}(2017)}]{Gross2017}%
  \BibitemOpen
  \bibfield  {author} {\bibinfo {author} {\bibfnamefont {C.}~\bibnamefont
  {Gross}}\ and\ \bibinfo {author} {\bibfnamefont {I.}~\bibnamefont {Bloch}},\
  }\href {\doibase 10.1126/science.aal3837} {\bibfield  {journal} {\bibinfo
  {journal} {Science}\ }\textbf {\bibinfo {volume} {357}},\ \bibinfo {pages}
  {995} (\bibinfo {year} {2017})}\BibitemShut {NoStop}%
\bibitem [{\citenamefont {Bloch}\ \emph {et~al.}(2008)\citenamefont {Bloch},
  \citenamefont {Dalibard},\ and\ \citenamefont {Zwerger}}]{Bloch2008r}%
  \BibitemOpen
  \bibfield  {author} {\bibinfo {author} {\bibfnamefont {I.}~\bibnamefont
  {Bloch}}, \bibinfo {author} {\bibfnamefont {J.}~\bibnamefont {Dalibard}}, \
  and\ \bibinfo {author} {\bibfnamefont {W.}~\bibnamefont {Zwerger}},\ }\href
  {\doibase 10.1103/RevModPhys.80.885} {\bibfield  {journal} {\bibinfo
  {journal} {Rev. Mod. Phys.}\ }\textbf {\bibinfo {volume} {80}},\ \bibinfo
  {pages} {885} (\bibinfo {year} {2008})}\BibitemShut {NoStop}%
\bibitem [{\citenamefont {Greif}\ \emph {et~al.}(2013)\citenamefont {Greif},
  \citenamefont {Uehlinger}, \citenamefont {Jotzu}, \citenamefont {Tarruell},\
  and\ \citenamefont {Esslinger}}]{Greif2013}%
  \BibitemOpen
  \bibfield  {author} {\bibinfo {author} {\bibfnamefont {D.}~\bibnamefont
  {Greif}}, \bibinfo {author} {\bibfnamefont {T.}~\bibnamefont {Uehlinger}},
  \bibinfo {author} {\bibfnamefont {G.}~\bibnamefont {Jotzu}}, \bibinfo
  {author} {\bibfnamefont {L.}~\bibnamefont {Tarruell}}, \ and\ \bibinfo
  {author} {\bibfnamefont {T.}~\bibnamefont {Esslinger}},\ }\href {\doibase
  10.1126/science.1236362} {\bibfield  {journal} {\bibinfo  {journal}
  {Science}\ }\textbf {\bibinfo {volume} {340}},\ \bibinfo {pages} {1307}
  (\bibinfo {year} {2013})}\BibitemShut {NoStop}%
\bibitem [{\citenamefont {Zeiher}\ \emph {et~al.}(2017)\citenamefont {Zeiher},
  \citenamefont {Choi}, \citenamefont {Rubio-Abadal}, \citenamefont {Pohl},
  \citenamefont {van Bijnen}, \citenamefont {Bloch},\ and\ \citenamefont
  {Gross}}]{Zeiher2017}%
  \BibitemOpen
  \bibfield  {author} {\bibinfo {author} {\bibfnamefont {J.}~\bibnamefont
  {Zeiher}}, \bibinfo {author} {\bibfnamefont {J.-y.}\ \bibnamefont {Choi}},
  \bibinfo {author} {\bibfnamefont {A.}~\bibnamefont {Rubio-Abadal}}, \bibinfo
  {author} {\bibfnamefont {T.}~\bibnamefont {Pohl}}, \bibinfo {author}
  {\bibfnamefont {R.}~\bibnamefont {van Bijnen}}, \bibinfo {author}
  {\bibfnamefont {I.}~\bibnamefont {Bloch}}, \ and\ \bibinfo {author}
  {\bibfnamefont {C.}~\bibnamefont {Gross}},\ }\href {\doibase
  10.1103/PhysRevX.7.041063} {\bibfield  {journal} {\bibinfo  {journal} {Phys.
  Rev. X}\ }\textbf {\bibinfo {volume} {7}},\ \bibinfo {pages} {041063}
  (\bibinfo {year} {2017})}\BibitemShut {NoStop}%
\bibitem [{\citenamefont {Bernien}\ \emph {et~al.}(2017)\citenamefont
  {Bernien}, \citenamefont {Schwartz}, \citenamefont {Keesling}, \citenamefont
  {Levine}, \citenamefont {Omran}, \citenamefont {Pichler}, \citenamefont
  {Choi}, \citenamefont {Zibrov}, \citenamefont {Endres}, \citenamefont
  {Greiner}, \citenamefont {Vuletić},\ and\ \citenamefont
  {Lukin}}]{Bernien2017}%
  \BibitemOpen
  \bibfield  {author} {\bibinfo {author} {\bibfnamefont {H.}~\bibnamefont
  {Bernien}}, \bibinfo {author} {\bibfnamefont {S.}~\bibnamefont {Schwartz}},
  \bibinfo {author} {\bibfnamefont {A.}~\bibnamefont {Keesling}}, \bibinfo
  {author} {\bibfnamefont {H.}~\bibnamefont {Levine}}, \bibinfo {author}
  {\bibfnamefont {A.}~\bibnamefont {Omran}}, \bibinfo {author} {\bibfnamefont
  {H.}~\bibnamefont {Pichler}}, \bibinfo {author} {\bibfnamefont
  {S.}~\bibnamefont {Choi}}, \bibinfo {author} {\bibfnamefont {A.~S.}\
  \bibnamefont {Zibrov}}, \bibinfo {author} {\bibfnamefont {M.}~\bibnamefont
  {Endres}}, \bibinfo {author} {\bibfnamefont {M.}~\bibnamefont {Greiner}},
  \bibinfo {author} {\bibfnamefont {V.}~\bibnamefont {Vuletić}}, \ and\
  \bibinfo {author} {\bibfnamefont {M.~D.}\ \bibnamefont {Lukin}},\ }\href
  {\doibase 10.1038/nature24622} {\bibfield  {journal} {\bibinfo  {journal}
  {Nature}\ }\textbf {\bibinfo {volume} {551}},\ \bibinfo {pages} {579}
  (\bibinfo {year} {2017})}\BibitemShut {NoStop}%
\bibitem [{\citenamefont {Barredo}\ \emph {et~al.}(2018)\citenamefont
  {Barredo}, \citenamefont {Lienhard}, \citenamefont {de~Leseleuc},
  \citenamefont {Lahaye},\ and\ \citenamefont {Browaeys}}]{Barredo2018}%
  \BibitemOpen
  \bibfield  {author} {\bibinfo {author} {\bibfnamefont {D.}~\bibnamefont
  {Barredo}}, \bibinfo {author} {\bibfnamefont {V.}~\bibnamefont {Lienhard}},
  \bibinfo {author} {\bibfnamefont {S.}~\bibnamefont {de~Leseleuc}}, \bibinfo
  {author} {\bibfnamefont {T.}~\bibnamefont {Lahaye}}, \ and\ \bibinfo {author}
  {\bibfnamefont {A.}~\bibnamefont {Browaeys}},\ }\href {\doibase
  10.1038/s41586-018-0450-2} {\bibfield  {journal} {\bibinfo  {journal}
  {Nature}\ }\textbf {\bibinfo {volume} {561}},\ \bibinfo {pages} {79}
  (\bibinfo {year} {2018})}\BibitemShut {NoStop}%
\bibitem [{\citenamefont {Guardado-Sanchez}\ \emph {et~al.}(2018)\citenamefont
  {Guardado-Sanchez}, \citenamefont {Brown}, \citenamefont {Mitra},
  \citenamefont {Devakul}, \citenamefont {Huse}, \citenamefont {Schau\ss{}},\
  and\ \citenamefont {Bakr}}]{Guardado2018}%
  \BibitemOpen
  \bibfield  {author} {\bibinfo {author} {\bibfnamefont {E.}~\bibnamefont
  {Guardado-Sanchez}}, \bibinfo {author} {\bibfnamefont {P.~T.}\ \bibnamefont
  {Brown}}, \bibinfo {author} {\bibfnamefont {D.}~\bibnamefont {Mitra}},
  \bibinfo {author} {\bibfnamefont {T.}~\bibnamefont {Devakul}}, \bibinfo
  {author} {\bibfnamefont {D.~A.}\ \bibnamefont {Huse}}, \bibinfo {author}
  {\bibfnamefont {P.}~\bibnamefont {Schau\ss{}}}, \ and\ \bibinfo {author}
  {\bibfnamefont {W.~S.}\ \bibnamefont {Bakr}},\ }\href {\doibase
  10.1103/PhysRevX.8.021069} {\bibfield  {journal} {\bibinfo  {journal} {Phys.
  Rev. X}\ }\textbf {\bibinfo {volume} {8}},\ \bibinfo {pages} {021069}
  (\bibinfo {year} {2018})}\BibitemShut {NoStop}%
\bibitem [{\citenamefont {Neyenhuis}\ \emph {et~al.}(2017)\citenamefont
  {Neyenhuis}, \citenamefont {Zhang}, \citenamefont {Hess}, \citenamefont
  {Smith}, \citenamefont {Lee}, \citenamefont {Richerme}, \citenamefont {Gong},
  \citenamefont {Gorshkov},\ and\ \citenamefont {Monroe}}]{Neyenhuise2017}%
  \BibitemOpen
  \bibfield  {author} {\bibinfo {author} {\bibfnamefont {B.}~\bibnamefont
  {Neyenhuis}}, \bibinfo {author} {\bibfnamefont {J.}~\bibnamefont {Zhang}},
  \bibinfo {author} {\bibfnamefont {P.~W.}\ \bibnamefont {Hess}}, \bibinfo
  {author} {\bibfnamefont {J.}~\bibnamefont {Smith}}, \bibinfo {author}
  {\bibfnamefont {A.~C.}\ \bibnamefont {Lee}}, \bibinfo {author} {\bibfnamefont
  {P.}~\bibnamefont {Richerme}}, \bibinfo {author} {\bibfnamefont {Z.-X.}\
  \bibnamefont {Gong}}, \bibinfo {author} {\bibfnamefont {A.~V.}\ \bibnamefont
  {Gorshkov}}, \ and\ \bibinfo {author} {\bibfnamefont {C.}~\bibnamefont
  {Monroe}},\ }\href {\doibase 10.1126/sciadv.1700672} {\bibfield  {journal}
  {\bibinfo  {journal} {Science Advances}\ }\textbf {\bibinfo {volume} {3}},\
  \bibinfo {pages} {e1700672} (\bibinfo {year} {2017})}\BibitemShut {NoStop}%
\bibitem [{\citenamefont {Blatt}\ and\ \citenamefont {Roos}(2012)}]{Blatt2012}%
  \BibitemOpen
  \bibfield  {author} {\bibinfo {author} {\bibfnamefont {R.}~\bibnamefont
  {Blatt}}\ and\ \bibinfo {author} {\bibfnamefont {C.~F.}\ \bibnamefont
  {Roos}},\ }\href {https://doi.org/10.1038/nphys2252} {\bibfield  {journal}
  {\bibinfo  {journal} {Nature Physics}\ }\textbf {\bibinfo {volume} {8}},\
  \bibinfo {pages} {277} (\bibinfo {year} {2012})}\BibitemShut {NoStop}%
\bibitem [{\citenamefont {Britton}\ \emph {et~al.}(2012)\citenamefont
  {Britton}, \citenamefont {Sawyer}, \citenamefont {Keith}, \citenamefont
  {Wang}, \citenamefont {Freericks}, \citenamefont {Uys}, \citenamefont
  {Biercuk},\ and\ \citenamefont {Bollinger}}]{Britton2012}%
  \BibitemOpen
  \bibfield  {author} {\bibinfo {author} {\bibfnamefont {J.~W.}\ \bibnamefont
  {Britton}}, \bibinfo {author} {\bibfnamefont {B.~C.}\ \bibnamefont {Sawyer}},
  \bibinfo {author} {\bibfnamefont {A.~C.}\ \bibnamefont {Keith}}, \bibinfo
  {author} {\bibfnamefont {C.~C.~J.}\ \bibnamefont {Wang}}, \bibinfo {author}
  {\bibfnamefont {J.~K.}\ \bibnamefont {Freericks}}, \bibinfo {author}
  {\bibfnamefont {H.}~\bibnamefont {Uys}}, \bibinfo {author} {\bibfnamefont
  {M.~J.}\ \bibnamefont {Biercuk}}, \ and\ \bibinfo {author} {\bibfnamefont
  {J.~J.}\ \bibnamefont {Bollinger}},\ }\href
  {https://doi.org/10.1038/nature10981} {\bibfield  {journal} {\bibinfo
  {journal} {Nature}\ }\textbf {\bibinfo {volume} {484}},\ \bibinfo {pages}
  {489} (\bibinfo {year} {2012})}\BibitemShut {NoStop}%
\bibitem [{\citenamefont {Yan}\ \emph {et~al.}(2013)\citenamefont {Yan},
  \citenamefont {Moses}, \citenamefont {Gadway}, \citenamefont {Covey},
  \citenamefont {Hazzard}, \citenamefont {Rey}, \citenamefont {Jin},\ and\
  \citenamefont {Ye}}]{Yan2013ood}%
  \BibitemOpen
  \bibfield  {author} {\bibinfo {author} {\bibfnamefont {B.}~\bibnamefont
  {Yan}}, \bibinfo {author} {\bibfnamefont {S.~A.}\ \bibnamefont {Moses}},
  \bibinfo {author} {\bibfnamefont {B.}~\bibnamefont {Gadway}}, \bibinfo
  {author} {\bibfnamefont {J.~P.}\ \bibnamefont {Covey}}, \bibinfo {author}
  {\bibfnamefont {K.~R.~A.}\ \bibnamefont {Hazzard}}, \bibinfo {author}
  {\bibfnamefont {A.~M.}\ \bibnamefont {Rey}}, \bibinfo {author} {\bibfnamefont
  {D.~S.}\ \bibnamefont {Jin}}, \ and\ \bibinfo {author} {\bibfnamefont
  {J.}~\bibnamefont {Ye}},\ }\href {\doibase 10.1038/nature12483} {\bibfield
  {journal} {\bibinfo  {journal} {Nature}\ }\textbf {\bibinfo {volume} {501}},\
  \bibinfo {pages} {521} (\bibinfo {year} {2013})}\BibitemShut {NoStop}%
\bibitem [{\citenamefont {Hazzard}\ \emph {et~al.}(2014)\citenamefont
  {Hazzard}, \citenamefont {Gadway}, \citenamefont {Foss-Feig}, \citenamefont
  {Yan}, \citenamefont {Moses}, \citenamefont {Covey}, \citenamefont {Yao},
  \citenamefont {Lukin}, \citenamefont {Ye}, \citenamefont {Jin},\ and\
  \citenamefont {Rey}}]{Hazzard2014}%
  \BibitemOpen
  \bibfield  {author} {\bibinfo {author} {\bibfnamefont {K.~R.~A.}\
  \bibnamefont {Hazzard}}, \bibinfo {author} {\bibfnamefont {B.}~\bibnamefont
  {Gadway}}, \bibinfo {author} {\bibfnamefont {M.}~\bibnamefont {Foss-Feig}},
  \bibinfo {author} {\bibfnamefont {B.}~\bibnamefont {Yan}}, \bibinfo {author}
  {\bibfnamefont {S.~A.}\ \bibnamefont {Moses}}, \bibinfo {author}
  {\bibfnamefont {J.~P.}\ \bibnamefont {Covey}}, \bibinfo {author}
  {\bibfnamefont {N.~Y.}\ \bibnamefont {Yao}}, \bibinfo {author} {\bibfnamefont
  {M.~D.}\ \bibnamefont {Lukin}}, \bibinfo {author} {\bibfnamefont
  {J.}~\bibnamefont {Ye}}, \bibinfo {author} {\bibfnamefont {D.~S.}\
  \bibnamefont {Jin}}, \ and\ \bibinfo {author} {\bibfnamefont {A.~M.}\
  \bibnamefont {Rey}},\ }\href {\doibase 10.1103/PhysRevLett.113.195302}
  {\bibfield  {journal} {\bibinfo  {journal} {Phys. Rev. Lett.}\ }\textbf
  {\bibinfo {volume} {113}},\ \bibinfo {pages} {195302} (\bibinfo {year}
  {2014})}\BibitemShut {NoStop}%
\bibitem [{\citenamefont {de~Paz}\ \emph {et~al.}(2013)\citenamefont {de~Paz},
  \citenamefont {Sharma}, \citenamefont {Chotia}, \citenamefont {Mar\'echal},
  \citenamefont {Huckans}, \citenamefont {Pedri}, \citenamefont {Santos},
  \citenamefont {Gorceix}, \citenamefont {Vernac},\ and\ \citenamefont
  {Laburthe-Tolra}}]{dePaz2013}%
  \BibitemOpen
  \bibfield  {author} {\bibinfo {author} {\bibfnamefont {A.}~\bibnamefont
  {de~Paz}}, \bibinfo {author} {\bibfnamefont {A.}~\bibnamefont {Sharma}},
  \bibinfo {author} {\bibfnamefont {A.}~\bibnamefont {Chotia}}, \bibinfo
  {author} {\bibfnamefont {E.}~\bibnamefont {Mar\'echal}}, \bibinfo {author}
  {\bibfnamefont {J.~H.}\ \bibnamefont {Huckans}}, \bibinfo {author}
  {\bibfnamefont {P.}~\bibnamefont {Pedri}}, \bibinfo {author} {\bibfnamefont
  {L.}~\bibnamefont {Santos}}, \bibinfo {author} {\bibfnamefont
  {O.}~\bibnamefont {Gorceix}}, \bibinfo {author} {\bibfnamefont
  {L.}~\bibnamefont {Vernac}}, \ and\ \bibinfo {author} {\bibfnamefont
  {B.}~\bibnamefont {Laburthe-Tolra}},\ }\href {\doibase
  10.1103/PhysRevLett.111.185305} {\bibfield  {journal} {\bibinfo  {journal}
  {Phys. Rev. Lett.}\ }\textbf {\bibinfo {volume} {111}},\ \bibinfo {pages}
  {185305} (\bibinfo {year} {2013})}\BibitemShut {NoStop}%
\bibitem [{\citenamefont {de~Paz}\ \emph {et~al.}(2016)\citenamefont {de~Paz},
  \citenamefont {Pedri}, \citenamefont {Sharma}, \citenamefont {Efremov},
  \citenamefont {Naylor}, \citenamefont {Gorceix}, \citenamefont {Mar\'echal},
  \citenamefont {Vernac},\ and\ \citenamefont {Laburthe-Tolra}}]{dePaz2016}%
  \BibitemOpen
  \bibfield  {author} {\bibinfo {author} {\bibfnamefont {A.}~\bibnamefont
  {de~Paz}}, \bibinfo {author} {\bibfnamefont {P.}~\bibnamefont {Pedri}},
  \bibinfo {author} {\bibfnamefont {A.}~\bibnamefont {Sharma}}, \bibinfo
  {author} {\bibfnamefont {M.}~\bibnamefont {Efremov}}, \bibinfo {author}
  {\bibfnamefont {B.}~\bibnamefont {Naylor}}, \bibinfo {author} {\bibfnamefont
  {O.}~\bibnamefont {Gorceix}}, \bibinfo {author} {\bibfnamefont
  {E.}~\bibnamefont {Mar\'echal}}, \bibinfo {author} {\bibfnamefont
  {L.}~\bibnamefont {Vernac}}, \ and\ \bibinfo {author} {\bibfnamefont
  {B.}~\bibnamefont {Laburthe-Tolra}},\ }\href {\doibase
  10.1103/PhysRevA.93.021603} {\bibfield  {journal} {\bibinfo  {journal} {Phys.
  Rev. A}\ }\textbf {\bibinfo {volume} {93}},\ \bibinfo {pages} {021603}
  (\bibinfo {year} {2016})}\BibitemShut {NoStop}%
\bibitem [{\citenamefont {Lepoutre}\ \emph {et~al.}(2018)\citenamefont
  {Lepoutre}, \citenamefont {Schachenmayer}, \citenamefont {Gabardos},
  \citenamefont {Zhu}, \citenamefont {Naylor}, \citenamefont {Marechal},
  \citenamefont {Gorceix}, \citenamefont {Rey}, \citenamefont {Vernac},\ and\
  \citenamefont {Laburthe-Tolra}}]{Lepoutre2018}%
  \BibitemOpen
  \bibfield  {author} {\bibinfo {author} {\bibfnamefont {S.}~\bibnamefont
  {Lepoutre}}, \bibinfo {author} {\bibfnamefont {J.}~\bibnamefont
  {Schachenmayer}}, \bibinfo {author} {\bibfnamefont {L.}~\bibnamefont
  {Gabardos}}, \bibinfo {author} {\bibfnamefont {B.}~\bibnamefont {Zhu}},
  \bibinfo {author} {\bibfnamefont {B.}~\bibnamefont {Naylor}}, \bibinfo
  {author} {\bibfnamefont {E.}~\bibnamefont {Marechal}}, \bibinfo {author}
  {\bibfnamefont {O.}~\bibnamefont {Gorceix}}, \bibinfo {author} {\bibfnamefont
  {A.~M.}\ \bibnamefont {Rey}}, \bibinfo {author} {\bibfnamefont
  {L.}~\bibnamefont {Vernac}}, \ and\ \bibinfo {author} {\bibfnamefont
  {B.}~\bibnamefont {Laburthe-Tolra}},\ }\href@noop {} {\bibfield  {journal}
  {\bibinfo  {journal} {arxiv}\ }\textbf {\bibinfo {volume} {1803.02628}}
  (\bibinfo {year} {2018})}\BibitemShut {NoStop}%
\bibitem [{\citenamefont {Stamper-Kurn}\ and\ \citenamefont
  {Ueda}(2013)}]{Stamper-Kurn2013}%
  \BibitemOpen
  \bibfield  {author} {\bibinfo {author} {\bibfnamefont {D.~M.}\ \bibnamefont
  {Stamper-Kurn}}\ and\ \bibinfo {author} {\bibfnamefont {M.}~\bibnamefont
  {Ueda}},\ }\href {\doibase 10.1103/RevModPhys.85.1191} {\bibfield  {journal}
  {\bibinfo  {journal} {Rev. Mod. Phys.}\ }\textbf {\bibinfo {volume} {85}},\
  \bibinfo {pages} {1191} (\bibinfo {year} {2013})}\BibitemShut {NoStop}%
\bibitem [{\citenamefont {Becher}\ \emph {et~al.}(2018)\citenamefont {Becher},
  \citenamefont {Baier}, \citenamefont {Aikawa}, \citenamefont {Lepers},
  \citenamefont {Wyart}, \citenamefont {Dulieu},\ and\ \citenamefont
  {Ferlaino}}]{Becher2017pol}%
  \BibitemOpen
  \bibfield  {author} {\bibinfo {author} {\bibfnamefont {J.~H.}\ \bibnamefont
  {Becher}}, \bibinfo {author} {\bibfnamefont {S.}~\bibnamefont {Baier}},
  \bibinfo {author} {\bibfnamefont {K.}~\bibnamefont {Aikawa}}, \bibinfo
  {author} {\bibfnamefont {M.}~\bibnamefont {Lepers}}, \bibinfo {author}
  {\bibfnamefont {J.-F.}\ \bibnamefont {Wyart}}, \bibinfo {author}
  {\bibfnamefont {O.}~\bibnamefont {Dulieu}}, \ and\ \bibinfo {author}
  {\bibfnamefont {F.}~\bibnamefont {Ferlaino}},\ }\href {\doibase
  10.1103/PhysRevA.97.012509} {\bibfield  {journal} {\bibinfo  {journal} {Phys.
  Rev. A}\ }\textbf {\bibinfo {volume} {97}},\ \bibinfo {pages} {012509}
  (\bibinfo {year} {2018})}\BibitemShut {NoStop}%
\bibitem [{sup()}]{suppmat}%
  \BibitemOpen
  \href@noop {} {}\bibinfo {note} {Materials, methods, and additional
  theoretical background are available as supplementary material on Science
  Online.}\BibitemShut {Stop}%
\bibitem [{\citenamefont {Auerbach}(1994)}]{Auerbach1994iea}%
  \BibitemOpen
  \bibfield  {author} {\bibinfo {author} {\bibfnamefont {A.}~\bibnamefont
  {Auerbach}},\ }\href@noop {} {\emph {\bibinfo {title} {Interacting Electrons
  and Quantum Magnetism}}}\ (\bibinfo  {publisher} {Springer-Verlag, New
  York},\ \bibinfo {year} {1994})\BibitemShut {NoStop}%
\bibitem [{\citenamefont {Dutta}\ \emph {et~al.}(2015)\citenamefont {Dutta},
  \citenamefont {Gajda}, \citenamefont {Hauke}, \citenamefont {Lewenstein},
  \citenamefont {L\"uhmann}, \citenamefont {Malomed}, \citenamefont
  {Sowi\'{n}ski},\ and\ \citenamefont {Zakrzewski}}]{Dutta2015}%
  \BibitemOpen
  \bibfield  {author} {\bibinfo {author} {\bibfnamefont {O.}~\bibnamefont
  {Dutta}}, \bibinfo {author} {\bibfnamefont {M.}~\bibnamefont {Gajda}},
  \bibinfo {author} {\bibfnamefont {P.}~\bibnamefont {Hauke}}, \bibinfo
  {author} {\bibfnamefont {M.}~\bibnamefont {Lewenstein}}, \bibinfo {author}
  {\bibfnamefont {D.-S.}\ \bibnamefont {L\"uhmann}}, \bibinfo {author}
  {\bibfnamefont {B.~A.}\ \bibnamefont {Malomed}}, \bibinfo {author}
  {\bibfnamefont {T.}~\bibnamefont {Sowi\'{n}ski}}, \ and\ \bibinfo {author}
  {\bibfnamefont {J.}~\bibnamefont {Zakrzewski}},\ }\href
  {http://stacks.iop.org/0034-4885/78/i=6/a=066001} {\bibfield  {journal}
  {\bibinfo  {journal} {Reports on Progress in Physics}\ }\textbf {\bibinfo
  {volume} {78}},\ \bibinfo {pages} {066001} (\bibinfo {year}
  {2015})}\BibitemShut {NoStop}%
\bibitem [{\citenamefont {Baier}\ \emph {et~al.}(2018)\citenamefont {Baier},
  \citenamefont {Petter}, \citenamefont {Becher}, \citenamefont {Patscheider},
  \citenamefont {Natale}, \citenamefont {Chomaz}, \citenamefont {Mark},\ and\
  \citenamefont {Ferlaino}}]{Baier2017sif}%
  \BibitemOpen
  \bibfield  {author} {\bibinfo {author} {\bibfnamefont {S.}~\bibnamefont
  {Baier}}, \bibinfo {author} {\bibfnamefont {D.}~\bibnamefont {Petter}},
  \bibinfo {author} {\bibfnamefont {J.~H.}\ \bibnamefont {Becher}}, \bibinfo
  {author} {\bibfnamefont {A.}~\bibnamefont {Patscheider}}, \bibinfo {author}
  {\bibfnamefont {G.}~\bibnamefont {Natale}}, \bibinfo {author} {\bibfnamefont
  {L.}~\bibnamefont {Chomaz}}, \bibinfo {author} {\bibfnamefont {M.~J.}\
  \bibnamefont {Mark}}, \ and\ \bibinfo {author} {\bibfnamefont
  {F.}~\bibnamefont {Ferlaino}},\ }\href {\doibase
  10.1103/PhysRevLett.121.093602} {\bibfield  {journal} {\bibinfo  {journal}
  {Phys. Rev. Lett.}\ }\textbf {\bibinfo {volume} {121}},\ \bibinfo {pages}
  {093602} (\bibinfo {year} {2018})}\BibitemShut {NoStop}%
\bibitem [{\citenamefont {Schachenmayer}\ \emph
  {et~al.}(2015{\natexlab{a}})\citenamefont {Schachenmayer}, \citenamefont
  {Pikovski},\ and\ \citenamefont {Rey}}]{Schachenmayer2015a}%
  \BibitemOpen
  \bibfield  {author} {\bibinfo {author} {\bibfnamefont {J.}~\bibnamefont
  {Schachenmayer}}, \bibinfo {author} {\bibfnamefont {A.}~\bibnamefont
  {Pikovski}}, \ and\ \bibinfo {author} {\bibfnamefont {A.~M.}\ \bibnamefont
  {Rey}},\ }\href {\doibase 10.1088/1367-2630/17/6/065009} {\bibfield
  {journal} {\bibinfo  {journal} {New Journal of Physics}\ }\textbf {\bibinfo
  {volume} {17}},\ \bibinfo {pages} {065009} (\bibinfo {year}
  {2015}{\natexlab{a}})}\BibitemShut {NoStop}%
\bibitem [{\citenamefont {Schachenmayer}\ \emph
  {et~al.}(2015{\natexlab{b}})\citenamefont {Schachenmayer}, \citenamefont
  {Pikovski},\ and\ \citenamefont {Rey}}]{Schachenmayer2015b}%
  \BibitemOpen
  \bibfield  {author} {\bibinfo {author} {\bibfnamefont {J.}~\bibnamefont
  {Schachenmayer}}, \bibinfo {author} {\bibfnamefont {A.}~\bibnamefont
  {Pikovski}}, \ and\ \bibinfo {author} {\bibfnamefont {A.~M.}\ \bibnamefont
  {Rey}},\ }\href {\doibase 10.1103/PhysRevX.5.011022} {\bibfield  {journal}
  {\bibinfo  {journal} {Phys. Rev. X}\ }\textbf {\bibinfo {volume} {5}},\
  \bibinfo {pages} {011022} (\bibinfo {year} {2015}{\natexlab{b}})}\BibitemShut
  {NoStop}%
\bibitem [{\citenamefont {Polkovnikov}(2010)}]{Polkovnikov2010}%
  \BibitemOpen
  \bibfield  {author} {\bibinfo {author} {\bibfnamefont {A.}~\bibnamefont
  {Polkovnikov}},\ }\href {\doibase https://doi.org/10.1016/j.aop.2010.02.006}
  {\bibfield  {journal} {\bibinfo  {journal} {Annals of Physics}\ }\textbf
  {\bibinfo {volume} {325}},\ \bibinfo {pages} {1790 } (\bibinfo {year}
  {2010})}\BibitemShut {NoStop}%
\bibitem [{Not()}]{Note1}%
  \BibitemOpen
  \href@noop {} {}\bibinfo {note} {We note that dissipationless dynamics is the
  most justified on this time scale as losses stay below 14\%.}\BibitemShut
  {Stop}%
\bibitem [{\citenamefont {Mancini}\ \emph {et~al.}(2015)\citenamefont
  {Mancini}, \citenamefont {Pagano}, \citenamefont {Cappellini}, \citenamefont
  {Livi}, \citenamefont {Rider}, \citenamefont {Catani}, \citenamefont {Sias},
  \citenamefont {Zoller}, \citenamefont {Inguscio}, \citenamefont {Dalmonte},\
  and\ \citenamefont {Fallani}}]{Mancini1510}%
  \BibitemOpen
  \bibfield  {author} {\bibinfo {author} {\bibfnamefont {M.}~\bibnamefont
  {Mancini}}, \bibinfo {author} {\bibfnamefont {G.}~\bibnamefont {Pagano}},
  \bibinfo {author} {\bibfnamefont {G.}~\bibnamefont {Cappellini}}, \bibinfo
  {author} {\bibfnamefont {L.}~\bibnamefont {Livi}}, \bibinfo {author}
  {\bibfnamefont {M.}~\bibnamefont {Rider}}, \bibinfo {author} {\bibfnamefont
  {J.}~\bibnamefont {Catani}}, \bibinfo {author} {\bibfnamefont
  {C.}~\bibnamefont {Sias}}, \bibinfo {author} {\bibfnamefont {P.}~\bibnamefont
  {Zoller}}, \bibinfo {author} {\bibfnamefont {M.}~\bibnamefont {Inguscio}},
  \bibinfo {author} {\bibfnamefont {M.}~\bibnamefont {Dalmonte}}, \ and\
  \bibinfo {author} {\bibfnamefont {L.}~\bibnamefont {Fallani}},\ }\href
  {\doibase 10.1126/science.aaa8736} {\bibfield  {journal} {\bibinfo  {journal}
  {Science}\ }\textbf {\bibinfo {volume} {349}},\ \bibinfo {pages} {1510}
  (\bibinfo {year} {2015})}\BibitemShut {NoStop}%
\bibitem [{\citenamefont {Lanyon}\ \emph {et~al.}(2008)\citenamefont {Lanyon},
  \citenamefont {Barbieri}, \citenamefont {Almeida}, \citenamefont {Jennewein},
  \citenamefont {Ralph}, \citenamefont {Resch}, \citenamefont {Pryde},
  \citenamefont {O'Brien}, \citenamefont {Gilchrist},\ and\ \citenamefont
  {White}}]{Lanyon2008}%
  \BibitemOpen
  \bibfield  {author} {\bibinfo {author} {\bibfnamefont {B.~P.}\ \bibnamefont
  {Lanyon}}, \bibinfo {author} {\bibfnamefont {M.}~\bibnamefont {Barbieri}},
  \bibinfo {author} {\bibfnamefont {M.~P.}\ \bibnamefont {Almeida}}, \bibinfo
  {author} {\bibfnamefont {T.}~\bibnamefont {Jennewein}}, \bibinfo {author}
  {\bibfnamefont {T.~C.}\ \bibnamefont {Ralph}}, \bibinfo {author}
  {\bibfnamefont {K.~J.}\ \bibnamefont {Resch}}, \bibinfo {author}
  {\bibfnamefont {G.~J.}\ \bibnamefont {Pryde}}, \bibinfo {author}
  {\bibfnamefont {J.~L.}\ \bibnamefont {O'Brien}}, \bibinfo {author}
  {\bibfnamefont {A.}~\bibnamefont {Gilchrist}}, \ and\ \bibinfo {author}
  {\bibfnamefont {A.~G.}\ \bibnamefont {White}},\ }\href
  {https://doi.org/10.1038/nphys1150} {\bibfield  {journal} {\bibinfo
  {journal} {Nature Physics}\ }\textbf {\bibinfo {volume} {5}},\ \bibinfo
  {pages} {134} (\bibinfo {year} {2008})}\BibitemShut {NoStop}%
\bibitem [{\citenamefont {Brion}\ \emph {et~al.}(2007)\citenamefont {Brion},
  \citenamefont {M\o{}lmer},\ and\ \citenamefont {Saffman}}]{Brion2007}%
  \BibitemOpen
  \bibfield  {author} {\bibinfo {author} {\bibfnamefont {E.}~\bibnamefont
  {Brion}}, \bibinfo {author} {\bibfnamefont {K.}~\bibnamefont {M\o{}lmer}}, \
  and\ \bibinfo {author} {\bibfnamefont {M.}~\bibnamefont {Saffman}},\ }\href
  {\doibase 10.1103/PhysRevLett.99.260501} {\bibfield  {journal} {\bibinfo
  {journal} {Phys. Rev. Lett.}\ }\textbf {\bibinfo {volume} {99}},\ \bibinfo
  {pages} {260501} (\bibinfo {year} {2007})}\BibitemShut {NoStop}%
\bibitem [{\citenamefont {Kues}\ \emph {et~al.}(2017)\citenamefont {Kues},
  \citenamefont {Reimer}, \citenamefont {Roztocki}, \citenamefont {Cort{\'e}s},
  \citenamefont {Sciara}, \citenamefont {Wetzel}, \citenamefont {Zhang},
  \citenamefont {Cino}, \citenamefont {Chu}, \citenamefont {Little},
  \citenamefont {Moss}, \citenamefont {Caspani}, \citenamefont {Aza{\~n}a},\
  and\ \citenamefont {Morandotti}}]{Kues2017}%
  \BibitemOpen
  \bibfield  {author} {\bibinfo {author} {\bibfnamefont {M.}~\bibnamefont
  {Kues}}, \bibinfo {author} {\bibfnamefont {C.}~\bibnamefont {Reimer}},
  \bibinfo {author} {\bibfnamefont {P.}~\bibnamefont {Roztocki}}, \bibinfo
  {author} {\bibfnamefont {L.~R.}\ \bibnamefont {Cort{\'e}s}}, \bibinfo
  {author} {\bibfnamefont {S.}~\bibnamefont {Sciara}}, \bibinfo {author}
  {\bibfnamefont {B.}~\bibnamefont {Wetzel}}, \bibinfo {author} {\bibfnamefont
  {Y.}~\bibnamefont {Zhang}}, \bibinfo {author} {\bibfnamefont
  {A.}~\bibnamefont {Cino}}, \bibinfo {author} {\bibfnamefont {S.~T.}\
  \bibnamefont {Chu}}, \bibinfo {author} {\bibfnamefont {B.~E.}\ \bibnamefont
  {Little}}, \bibinfo {author} {\bibfnamefont {D.~J.}\ \bibnamefont {Moss}},
  \bibinfo {author} {\bibfnamefont {L.}~\bibnamefont {Caspani}}, \bibinfo
  {author} {\bibfnamefont {J.}~\bibnamefont {Aza{\~n}a}}, \ and\ \bibinfo
  {author} {\bibfnamefont {R.}~\bibnamefont {Morandotti}},\ }\href
  {https://doi.org/10.1038/nature22986} {\bibfield  {journal} {\bibinfo
  {journal} {Nature}\ }\textbf {\bibinfo {volume} {546}},\ \bibinfo {pages}
  {622} (\bibinfo {year} {2017})}\BibitemShut {NoStop}%
\bibitem [{\citenamefont {Aikawa}\ \emph {et~al.}(2014)\citenamefont {Aikawa},
  \citenamefont {Frisch}, \citenamefont {Mark}, \citenamefont {Baier},
  \citenamefont {Grimm},\ and\ \citenamefont {Ferlaino}}]{Aikawa2014rfd}%
  \BibitemOpen
  \bibfield  {author} {\bibinfo {author} {\bibfnamefont {K.}~\bibnamefont
  {Aikawa}}, \bibinfo {author} {\bibfnamefont {A.}~\bibnamefont {Frisch}},
  \bibinfo {author} {\bibfnamefont {M.}~\bibnamefont {Mark}}, \bibinfo {author}
  {\bibfnamefont {S.}~\bibnamefont {Baier}}, \bibinfo {author} {\bibfnamefont
  {R.}~\bibnamefont {Grimm}}, \ and\ \bibinfo {author} {\bibfnamefont
  {F.}~\bibnamefont {Ferlaino}},\ }\href {\doibase
  10.1103/PhysRevLett.112.010404} {\bibfield  {journal} {\bibinfo  {journal}
  {Phys. Rev. Lett.}\ }\textbf {\bibinfo {volume} {112}},\ \bibinfo {pages}
  {010404} (\bibinfo {year} {2014})}\BibitemShut {NoStop}%
\bibitem [{\citenamefont {Baier}\ \emph {et~al.}(2016)\citenamefont {Baier},
  \citenamefont {Mark}, \citenamefont {Petter}, \citenamefont {Aikawa},
  \citenamefont {Chomaz}, \citenamefont {Cai}, \citenamefont {Baranov},
  \citenamefont {Zoller},\ and\ \citenamefont {Ferlaino}}]{Baier2016ebh}%
  \BibitemOpen
  \bibfield  {author} {\bibinfo {author} {\bibfnamefont {S.}~\bibnamefont
  {Baier}}, \bibinfo {author} {\bibfnamefont {M.~J.}\ \bibnamefont {Mark}},
  \bibinfo {author} {\bibfnamefont {D.}~\bibnamefont {Petter}}, \bibinfo
  {author} {\bibfnamefont {K.}~\bibnamefont {Aikawa}}, \bibinfo {author}
  {\bibfnamefont {L.}~\bibnamefont {Chomaz}}, \bibinfo {author} {\bibfnamefont
  {Z.}~\bibnamefont {Cai}}, \bibinfo {author} {\bibfnamefont {M.}~\bibnamefont
  {Baranov}}, \bibinfo {author} {\bibfnamefont {P.}~\bibnamefont {Zoller}}, \
  and\ \bibinfo {author} {\bibfnamefont {F.}~\bibnamefont {Ferlaino}},\ }\href
  {\doibase 10.1126/science.aac9812} {\bibfield  {journal} {\bibinfo  {journal}
  {Science}\ }\textbf {\bibinfo {volume} {352}},\ \bibinfo {pages} {201}
  (\bibinfo {year} {2016})}\BibitemShut {NoStop}%
\bibitem [{\citenamefont {Bertlmann}\ and\ \citenamefont
  {Krammer}(2008)}]{gmm2008}%
  \BibitemOpen
  \bibfield  {author} {\bibinfo {author} {\bibfnamefont {R.~A.}\ \bibnamefont
  {Bertlmann}}\ and\ \bibinfo {author} {\bibfnamefont {P.}~\bibnamefont
  {Krammer}},\ }\href {\doibase 10.1088/1751-8113/41/23/235303} {\bibfield
  {journal} {\bibinfo  {journal} {Journal of Physics A: Mathematical and
  Theoretical}\ }\textbf {\bibinfo {volume} {41}},\ \bibinfo {pages} {235303}
  (\bibinfo {year} {2008})}\BibitemShut {NoStop}%
\end{thebibliography}%

\setcounter{equation}{0}
\setcounter{figure}{0}
\setcounter{table}{0}
\setcounter{page}{1}

\renewcommand{\theequation}{S\arabic{equation}}
\renewcommand{\thefigure}{S\arabic{figure}}
\renewcommand{\thesection}{S\arabic{section}}

\section*{Supplementary Materials}
\section{Experimental setup and lattice loading}
In our experiment we start with a degenerate Fermi gas of about $2.4\times 10^4$ ${}^{167}$Er atoms in the lowest spin state $|F=19/2, m_F=\text{--}19/2\rangle=\mnineteen$ and a temperature of $T \approx 0.3\,T_\text{F}$~\cite{Aikawa2014rfd, Baier2017sif}. The atoms are confined in a crossed optical dipole trap (ODT) and the trap frequencies are $(\nu_{\perp},\nu_{\parallel},\nu_{z})=(63(1),36(2),137(1))\,\rm Hz$, where $\nu_{\perp}$ ($\nu_\parallel$) are the trap frequencies perpendicular to (along) the horizontal ODT and $\nu_{z}$ is measured along the vertical direction defined by gravity. We load the atomic sample adiabatically into a 3D lattice that is created by two retro-reflected laser beams at $532\,$nm in the x-y plane and one retro-reflected laser beam at $1064\,$nm nearly along the z direction, defined by gravity and orthogonal to the x-y plane. Note, that due to a small angle of about $10^\circ$ between the vertical lattice beam and the z axis we obtain a 3D-lattice, slightly deviating from an ideal situation of a rectangular unit cell and our parallelepipedic cell has the unit lattice distances of $d_{x}\,=\,272\,{\rm nm}$, $d_{y}\,=\,266\,{\rm nm}$, and $d_{z}\,=\,544\,{\rm nm}$. The lattice geometry is similar to the one used in our previous works~\cite{Baier2016ebh,Baier2017sif}. We ramp up the lattice beams in $150\,$ms to their final power and switch off the ODT subsequently in $10\,$ms and wait for $500\,$ms to eliminate residual atoms in higher bands~\cite{Baier2017sif}. For our typical lattice depths used in the measurements reported here of $(s_{x},s_{y},s_{z})\,=\,(20,20,80)$, where $s_i$ with $i\in {x,y,z}$ is given in the respective recoil energy $\mathrm{E}_{\mathrm{R},i}$ with $E_{\text{R};x,y} = h \times 4.2\, \rm kHz$ and $E_{\text{R};z} = h \times 1.05\, \rm kHz$, the atoms can be considered pinned on single lattice sites with low tunneling rates $J_{x,y} = h \times 10.5\, \rm Hz$ and $J_{z} = h \times 1\, \rm mHz$.

\section{State preparation and detection efficiency}
\begin{figure}[t!]
	\centering
	\includegraphics[width=1\linewidth]{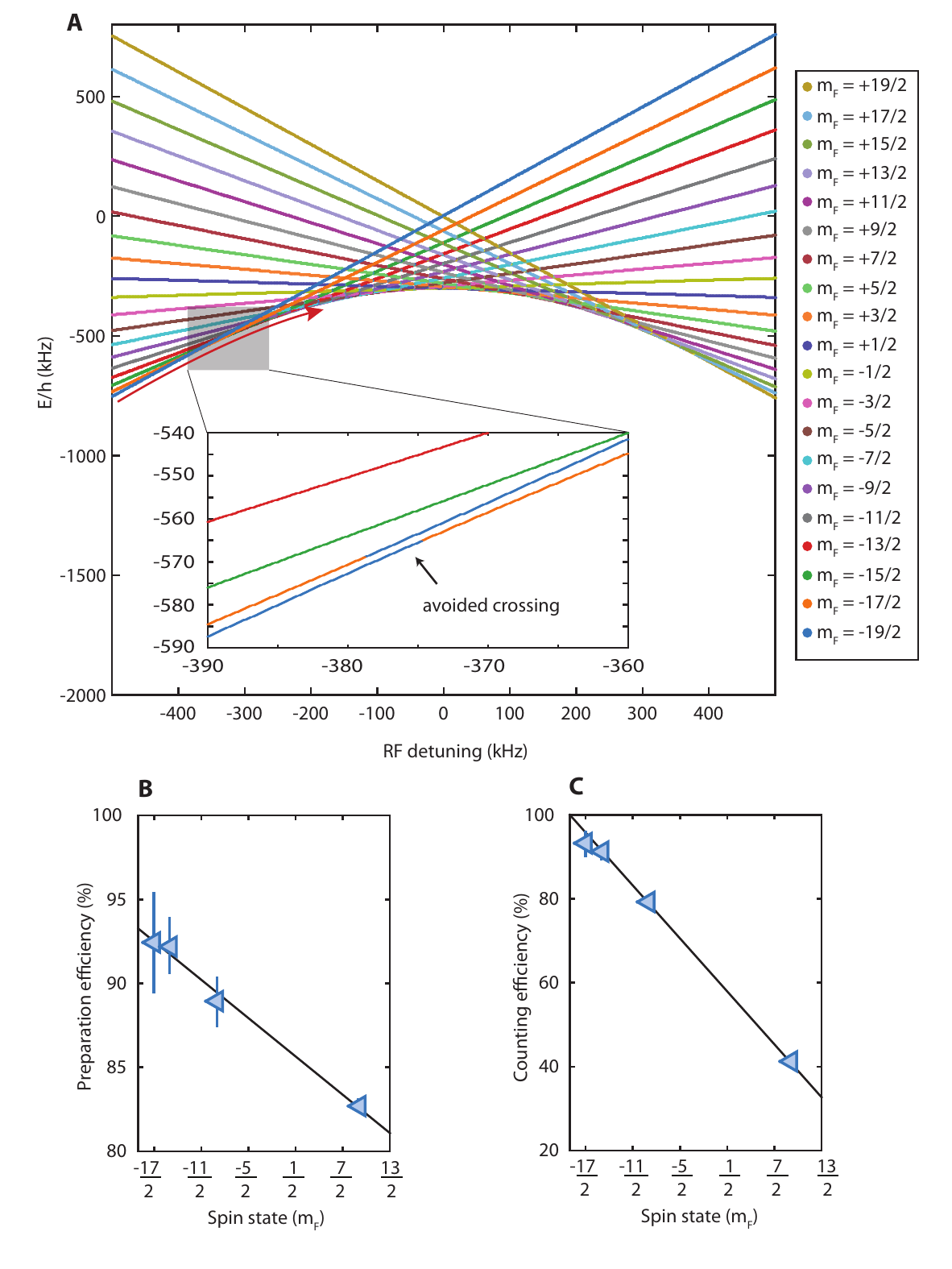}
    \caption{State preparation and detection efficiency. (A) Energy levels of the ground state hyperfine manifold in the dressed-state picture in dependence of the detuning between the applied RF-frequency and the atomic resonance condition for the $\mone$ to the $\pone$ hyperfine levels. The solid red arrow exemplary shows the RAP for preparation of atoms into the $|m_F^0 \rangle = \mseven$ state. The insets show a zoom of one avoided crossing. (B-C) Spin-preparation and atom-counting efficiency measured for $\mseventeen$, $\mfifteen$, $\mnine$, and $\pnine$. The obtained values are interpolated linearly assuming a linear dependence on the $m_F^0$ state.}
	\label{fig:S1}
\end{figure}

To prepare the atoms in the desired spin state, after loading them into the lattice, we use a technique based on a rapid-adiabatic passage (RAP). During the full preparation procedure, the optical lattice operates as a protection shield to avoid atom loss and heating due to the large density of Feshbach resonances~\cite{Baier2017sif}. To reach a reliable preparation with high fidelity it is necessary that the change in the energy difference between subsequent neighboring spin states is sufficiently large. Therefore, we ramp the magnetic field in $40\,\rm ms$ to $40.6\,\rm G$ to get a large enough differential quadratic Zeeman shift, which is on the order of $\approx h \times 40\,$kHz. After the magnetic field ramp we wait for $80\,\rm ms$ to allow the latter to stabilize before performing the RAP procedure. The follow up RAP protocol depends on the target state. For example, to transfer the atoms from $\mnineteen$  into the $\mseven$ state, we apply a radio-frequency (RF) pulse at $41.31\,\rm MHz$ and reduce the magnetic field with a linear ramp, e.\,g.\,by $500\,\rm mG$ in $42\,\rm ms$. The variation of the magnetic field is analogous to the more conventional scheme where the frequency of the RF is varied (see Fig.\,\ref{fig:S1}A). For the preparation of higher (lower) spin states we perform a larger (smaller) reduction of the magnetic field on a longer (shorter) timescale. After the RAP ramp we switch off the RF field and ramp the magnetic field in $10\,\rm ms$ to $B = 3.99\,\rm G$. Here we wait again for $100\,$ms to allow the magnetic field to stabilize. During the ramp up and down to $40\,$G of the magnetic field we loose $25(2)\,\%$ of the atoms. We attribute this loss mainly to the dense Feshbach spectra that we are crossing when ramping the magnetic field. The exact loss mechanism has not been yet identified, constituting a topic of interest for latter investigation. At $3.99\,$G, before switching on the spin dynamics, about $1.7\times 10^4$ atoms remain in the lattice. The losses affect the lattice filling at which the spin dynamics occur. Our simulations account for this initially reduced filling; see Sec.\,S8.

Additionally to the losses due to the magnetic field ramps, we also observe losses caused by the RAP itself. To quantify the preparation efficiency, i.\,e\,. the loss of atoms due to the spin preparation via RAP as a function of the target $m_F$ state, we perform additional measurements where we either do not perform a RAP or where we add an inverse RAP to transfer all atoms back into the $\mnineteen$ state. By comparing the atom numbers from measurements without RAP and with double-RAP and assuming that the loss process is symmetric, we derive the preparation efficiency as plotted in Fig.\,\ref{fig:S1}B. We also account for the difference in the counting efficiency of the individual spin states, which arises from different effective scattering cross sections for the imaging light. Here we compare the measured atom number in a target $m_F$ state to the expected atom number taking into account the previously determined preparation efficiency as discussed above and the atom number without RAP; see Fig.\,\ref{fig:S1}C.

The counting and preparation efficiencies are determined for the $\mseventeen$, $\mfifteen$, $\mnine$, and $\pnine$ states and interpolated assuming a linear dependency of these efficiencies on $m_F$ (see Fig.\,\ref{fig:S1}B,C). We estimate the preparation efficiency of the respective m$_F$ state to be $0.86(1) - 0.008(1)~\times~\text{m}_F$. We attribute the lower preparation efficiency for higher spin states to the larger number of avoided crossing between spin states that come into play during the RAP procedure. Overall we expect that the lattice filling over the whole sample, taking into account the losses due to magnetic field ramping and spin state preparation, reduces from close to unity down to a value between $0.6$ and $0.7$; see also Tab.\,S1 -- S2.

\section{Quench protocol and detection sequence}
\begin{figure}[t!]
	\centering
	\includegraphics[width=0.95\linewidth]{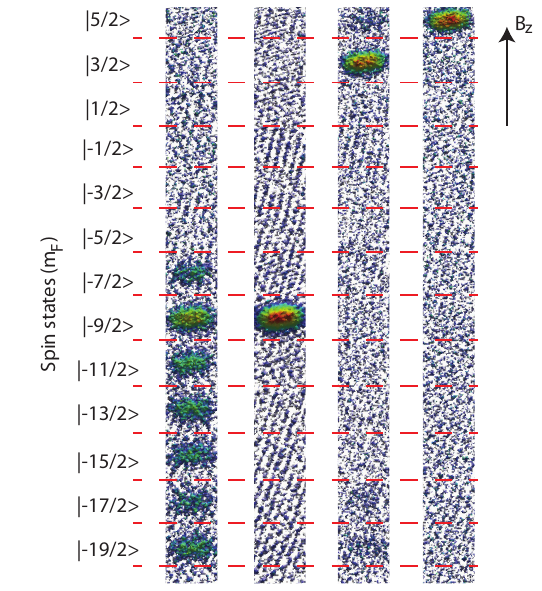}
    \caption{Spin resolved imaging. Absorption images for a non adjusted RAP and for the preparation of $\mnine$, $\pthree$, and $\pfive$. Whereas for the $\mnine$ and $\pfive$ case no residual atoms in other spin states are visible, for the $\pthree$ case we observe a small amount of residual atoms in other spin states due to a non perfect preparation. }
	\label{fig:S2}
\end{figure}

In our experiment we exploit both, the light and the magnetic shifts of the energies of each spin state to reach a resonant condition where the energy difference between neighboring spin states is cancelled and therefore spin changing dynamics preserving the total magnetization become energetically allowed. In particular we exploit the tensorial light shift of the spin states energies~\cite{Becher2017pol} $$U_t = \frac{3m_F^2 - F(F+1)}{F(2F-1)}\frac{3 \cos ^2 \theta _p -1}{2} \alpha _t (\omega),$$ present in atomic erbium to initialize the dynamic evolution of the spin population. The tensorial light shift depends quadratically on the $m_F$ state as well as on the angle $\theta _p$ between the magnetic field axis and the axis of polarization of the light. Here, $\alpha _t$ refers to the tensorial polarizability coefficient and $\omega$ to the light frequency. After the preparation of the respective spin state we start all our measurements at $B=3.99\, \rm G$, pointing in the $z$ direction. However, to reach the resonance condition we use two slightly different quench protocols for the measurement sets with fixed $\Theta = 0^\circ$ for the different initial spin state and for the sets of measurements where $|m_F \rangle =\mseventeen$ and $\Theta \in (0^\circ, 80^\circ)$. The measurement sequences differ on the one hand by the way we jump on resonance to initialize the spin dynamics and on the other hand by shining in an additional laser beam of wavelength $1064\, \rm nm$ and power of $7\, \rm W$. This additional light is necessary because changing $\Theta$ reduces simultaneously $\theta _p$ resulting in a smaller tensorial light shift and therefore in a shift of the resonance position to lower magnetic field values. For large $\Theta$ the light shift of the lattice beams is smaller and therefore the resonance is very close to $0\,$G which we want to avoid to prevent spin relaxation processes.
For the sets of measurements where we keep $\Theta = 0^\circ$ but vary the initial $m_F^0$ state we quench the magnetic field directly after the preparation, from $3.99\, \rm G$ to resonance. In contrast we use a different approach for the measurements where $\Theta$ is varied. After the preparation we ramp in $10\,\rm ms$ the additional laser beam to $7\,\rm W$. Due to the reduced speed of our magnetic field coils in $x$ and $y$ direction we first rotate the magnetic field such that the transverse components $B_{\rm{x}}$ and $B_{\rm{y}}$ are already at their target values while keeping an additional offset of $2\,$G in the  $z$ direction. The quench to resonance is then done using only the coils for the magnetic field in the $z$ direction. The additional offset field of $2\,$G is large enough to suppress dynamics. We measure the evolution of the magnetic field by performing RF spectroscopy and find that for both quench procedures the magnetic field evolves exponentially towards its quench value with $1/e$ decay times of $1.4\,\rm ms$ and $1.2\,\rm ms$, respectively.
After holding on resonance for a certain time we quench the magnetic field back to $3.99\, \rm G$ and we rotate the latter back to $\Theta = 0^\circ$. After a waiting time of $50 \, \rm ms$ we perform a band-mapping measurement combined with a Stern-Gerlach technique, i.e. we ramp the lattice down in $1\, \rm ms$ and apply a magnetic field gradient that is large enough to separate the individual spin states after a time of flight (TOF) of $15\,\rm ms$. This allows us to image the first Brillouin-zone for the different spin states. During TOF the magnetic field is rotated towards the imaging axis. We typically record the population of the initially prepared $\mfstate$, of its four neighbors, and of $\mnineteen$ by summing the 2D atomic density over a region of interest. Figure~\ref{fig:S2} shows examples of the imaging of different spin states for the cases of a non adjusted RAP as well as for the preparation of the atoms in $\mnine$, $\pthree$, and $\pfive$. In the case of $\pthree$ residual atoms in $\mnineteen$, $\mseventeen$, and $\pfive$ are visible due to a non perfect preparation.

\section{Lifetime and losses in the lattice}
Off-resonance, i.\,e.\, at a magnetic field of $B=3.99\,$G, we measure the lifetime of the prepared spin state to be on the order of a few seconds, being slightly shorter for higher spin states. Note that, here, we do not observe any population growing in the neighboring spin states. Differently, for the measurements on resonance we observe a faster loss happening on the timescale of the first $20$--$30\,$ms followed by loss at lower speed for the remaining atoms. We fit an exponential decay to extract the atoms loss and change in filling over the timescale that we use to extract $V_{\rm{eff}}$, $t_{\rm{fit}}$, (see S8) as well as over the full $100\,$ms of the dynamics reported in the main text (Fig.\,\ref{fig:2}-\ref{fig:3}). Table~S1 gives the corresponding numbers for the sets of data for the different initial $m_F^0$ states. During the fitting timescale we observe atoms loss on the order of $5$--$10$\%. This atom loss can be converted into a change of the effective filling of the lattice compared to the state obtained after the lattice loading giving a minimum filling of $\nu=0.58$ for the $|m_F^0 \rangle =\pone$ case. For longer timescales larger losses in the range between $10-35$\% are observed. In general, we note that the amount of loss depends on the initial $m_F^0$ state, resulting larger for the central $\mfstate^0$ states. Similar numbers are obtained for the sets of measurement where we vary $\Theta$ (see Tab.\,S2). The exact mechanism leading to these losses is not yet understood and will be the topic of future studies. Thanks to their limited importance over the early time dynamics, we here compare our results to theoretical prediction without losses; see S~8. A proper description of the long time dynamics will certainly require to account and understand such effects.

\begin{table*}
    \centering
    \caption{Percentage of lost atoms and extracted, effective filling fraction $\nu$  for different $|m_F \rangle$ states at $t=t_{\rm{fit}}$ and $t=100\,$ms.}
    \label{tab:S1}
    \begin{tabular}{c|c|c|c|c|c|c}
        $m_F^0$ & $t_{\rm{fit}}$ (ms) & $N_{\rm{loss}}(t_{\rm{fit}})$ (\%) & $\nu$ $(0)$ & $\nu$ $(t_{\rm{fit}})$ & $N_{\rm{loss}}(100\,$ms) (\%) & $\nu$ $(100\,$ms) \\
    \hline    
        $-\frac{17}{2}$ & $34.2$ & $1.8$ & $0.7$ & $0.68$ & $5.3$ & $0.66$ \\
        $-\frac{13}{2}$ & $15.7$ & $7.2$ & $0.69$ & $0.64$ & $19.6$ & $0.55$ \\
        $-\frac{9}{2}$ & $11.3$ & $8.7$ & $0.67$ & $0.62$ & $25.1$ & $0.51$\\
        $-\frac{5}{2}$ & $9.6$ & $13.7$ & $0.66$ & $0.58$ & $27.7$ & $0.48$ \\
        $-\frac{1}{2}$ & $9.0$ & $12.2$ & $0.65$ & $0.57$ & $36.1$ & $0.42$ \\
        $\frac{1}{2}$ & $9.0$ & $13.5$ & $0.65$ & $0.56$ & $36.7$ & $0.41$ \\
        $\frac{3}{2}$ & $9.2$ & $9.2$ & $0.64$ & $0.58$ & $34.0$ & $0.43$ \\
        $\frac{9}{2}$ & $11.3$ & $6.7$ & $0.63$ & $0.59$ & $23.1$ & $0.49$ \\
        $\frac{13}{2}$ & $15.7$ & $4.9$ & $0.62$ & $0.59$ & $21.9$ & $0.48$ \\
    \end{tabular}
\end{table*}

\begin{table*}
    \centering
    \caption{Percentage of lost atoms and extracted effective filling fraction $\nu$ for different $\Theta$ at $t=t_{\rm{fit}}$ and $t=100\,$ms.}
    \label{tab:S2}
    \begin{tabular}{c|c|c|c|c|c|c}
        $\Theta$ ($^\circ$) & $t_{\rm{fit}}$ (ms) & $N_{\rm{loss}}(t_{\rm{fit}})$ (\%) & $\nu$ $(0)$ & $\nu$ $(t_{\rm{fit}})$ & $N_{\rm{loss}}(100\,$ms) (\%) & $\nu$ $(100\,$ms) \\
    \hline    
        $0$ & $26.8$ &  $13.8$ & $0.7$ & $0.60$ & $30.1$ & $0.49$ \\
        $10$ & $30.1$ & $11.9$ & $0.7$ & $0.61$ & $25.6$ & $0.52$ \\
        $20$ & $36.7$ & $6.9$ & $0.7$ & $0.65$ & $18.2$ & $0.57$ \\
        $30$ & $47.0$ & $8.6$ & $0.7$ & $0.64$ & $17.8$ & $0.57$ \\
        $35$ & $52.0$ & $6.7$ & $0.7$ & $0.65$ & $11.4$ & $0.62$ \\
        $40$ & $53.2$ & $7.2$ & $0.7$ & $0.65$ & $13.2$ & $0.60$ \\
        $50$ & $46.1$ & $12$ & $0.7$ & $0.61$ & $17.2$ & $0.58$ \\
        $60$ & $37.0$ & $8.9$ & $0.7$ & $0.63$ & $19.2$ & $0.56$ \\
        $80$ & $30.0$ & $10.4$ & $0.7$ & $0.62$ & $19.3$ & $0.56$ \\
    \end{tabular}
\end{table*}

\section{Experimental uncertainties and inhomogeneities}
Ideally, all atoms in the sample experience the same linear and quadratic Zeeman shift and the same quadratic light shift. However, in the experiment inhomogeneities from the magnetic field and light intensities lead to a spatial dependence of those shifts. An upper bound of the variation of Zeeman shifts can be deduced from RF-spectroscopy measurements done with bosonic erbium. From the width of the RF-resonance ($\approx 500\,$Hz) and the size of the cloud ($\approx 15\, \rm{\mu}$m) we estimate a maximum magnetic field gradient of $\leq 230\,$mG/cm, assuming the gradient as main broadening mechanism for the resonance width, neglecting magnetic field noise and Fourier broadening. This translates into a differential linear Zeeman shift of $\leq h \times 6\,$Hz between adjacent lattice sites in the horizontal x-y plane and $\leq h \times 12\,$Hz between adjacent planes along the z-direction. Together with the magnetic field values used in the spin dynamic experiments, the variation of the quadratic Zeeman shift is negligible compared to other inhomogeneities ($\leq h \times 0.1\,$Hz). 
The inhomogeneity of the quadratic light shifts can be estimated by considering the shape of the lattice light beams (Gaussian beams with waists of about $(w_x,w_y,w_z) = (160,160,300)\, \rm{\mu}$m) and the resonance condition of the magnetic field, translated into a quadratic Zeeman shift of $h \times 71(1)\, \rm Hz$. This considerations can be used to obtain an estimation for a site dependent light shift compared to the center of the atomic sample. If we take a possible displacement of the atoms by $\leq 10\rm \mu m$ in all directions, from the center of the lattice to the center of the beams, into account, we can estimate an upper bound for the light shift of $\delta_i^T \leq h \times 2\,$Hz at $20$ lattice sites away from the center along the $y$ direction.

\section{Spin Hamiltonian} \label{genh} 
The experiment operates in a deep lattice regime, where tunneling is suppressed. At the achieved initial conditions, the $^{167}$Er atoms are restricted to occupy the lowest lattice band, and Fermi statistics prevents more than one atom per lattice site. In the presence of a magnetic field strong enough to generate Zeeman splittings larger than nearest-neighbor dipolar interactions, only those processes that conserve the total magnetization are energetically allowed~\cite{dePaz2016}. Under these considerations, the dynamics is described by the following secular Hamiltonian:
\begin{eqnarray}
    \hat H &=&\sum_i\delta_{ i} (\hat F_{ i} ^z)^2+\sum_{ i} B_{ i} \hat F_{ i} ^z \nonumber \\ && +\frac{1}{2}\sum_{i,j\neq i}V_{{ i} ,{ j}}[\hat F_{ i}^z\hat F_{ j}^z-\frac{1}{4}(\hat F_{ i}^+\hat F_{ j}^-+h.c)].\label{eq:H}
\end{eqnarray}
Here the operators $F_{ i}^{z,\pm}$ are spin $19/2$ angular momentum operators acting on lattice site $i$. The first two terms account for the site-dependent quadratic and linear shifts respectively, where $\delta_{ i}$  includes both Zeeman terms and tensorial light shifts as discussed in the main text. $B_{i}=B+\Delta B_i$ denotes the linear Zeeman shift at site $i$. While the constant and uniform contribution, $B$, commutes with all other terms, thus can be rotated out, the spatially varying contribution, $\Delta B_i$, is relatively small in the experiment but still is accounted for in the theory calculations.
The last term is the long-range dipolar interaction between atoms in different sites, with
\begin{eqnarray}
  V_{{\bf i},{\bf j}}&\equiv&V_{dd} d_y^3 \frac{1-3\cos^2\theta_{{ i},{ j}}}{|{\bf r}_{i}-{\bf r}_{ j}|^3},
\end{eqnarray}
where $\theta_{ij}$ is the angle between the dipolar orientation set by an external magnetic field and the inter-particle spacing ${\bf r}_{ i}-{\bf r}_{ j} $.  $V_{dd}\,\approx\,\frac{\mu_0g_F^2\mu_B^2}{4\pi d_y^3}$ corresponds to  the interaction strength between two atoms, $i$ and $j$, separated by the smallest lattice constant $|{\bf r}_{ i}-{\bf r}_{ j}|= d_y=266$\,nm   and  forming an angle  $\theta_{{ i},{ j}}=\pi/2$ with the quantization axis. Here  $g_F \approx 0.735$ is the Lande g-factor for Er atoms, $\mu_0$  is  the magnetic
permeability of vacuum and $\mu_B$ is the Bohr magneton. We compute $V_{dd}$ from 
\begin{eqnarray}
  V_{dd}&=&\frac{\mu_0(\mu_Bg_F)^2}{4\pi}\nonumber \\ && \times \int d^3{\bf r}d^3{\bf r}'\frac{1-3\cos^2\theta_{rr'}}{|{\bf r}-{\bf r}'|^3}|\phi_{ i}({\bf r})|^2|\phi_{ j}({\bf r}')|^2, \label{eq:Vdip}
\end{eqnarray}
where $\phi_{i}({ r})$ denotes the lowest band Wannier function centered at lattice site ${ i}$. For the experimental lattice depths $(s_x,s_y,s_z)=(20,20,80)$ in units of the corresponding recoil energies,  $V_{dd}$ is estimated to be $h \times 0.336\,$Hz.

\section{The GDTWA method} 
To account for quantum many-body effects during the  dynamics generated by   long-range dipolar interactions in these complex macroscopic spin $F=19/2$ 3D lattice array, we apply the so called Generalize Discrete Truncated Wigner Approximation (GDTWA) first introduced in Ref.~\cite{Lepoutre2018}. The underlying idea of the method is to supplement the mean field dynamics of a spin $F$ system with appropriate sampling over the initial conditions in order  to quantitatively account for the build up of quantum correlations. For a spin-F atom $i$ with $\mathcal{N}=2F+1$ spin states, its density matrix $\hat \rho_i$ consists of $\mathcal{D}=\mathcal{N}\times\mathcal{N}$ elements. Correspondingly, we can define $\mathcal{D}$ Hermitian operators, $\Lambda^i_{\mu}$, with $\mu=1,... D$, using the generalized Gell-Mann matrices (GGM) and the identity matrix~\cite{gmm2008}:
\begin{eqnarray}
    \Lambda^i_{\mu=1,...\mathcal{N}(\mathcal{N}-1)/2}&=&\frac{1}{\sqrt{2}}(\ket{\beta}\bra{\alpha}+h.c.), 
\end{eqnarray}
for $\alpha>\beta$, $1\le \alpha$,$\beta\le \mathcal{N}$,
\begin{eqnarray}
    \Lambda^i_{\mu=\mathcal{N}(\mathcal{N}-1)/2+1,...\mathcal{N}(\mathcal{N}-1)}&=&\frac{1}{\sqrt{2}i}(\ket{\beta}\bra{\alpha}-h.c.),
\end{eqnarray}
for $\alpha>\beta$, $1\le \alpha$,$\beta\le \mathcal{N}$,
\begin{eqnarray}
    \Lambda^i_{\mu=\mathcal{N}(\mathcal{N}-1)+1,...\mathcal{N}^2-1}&=& \frac{1}{\sqrt{\alpha(\alpha+1)}} \nonumber \\ &&\times (\sum_{\beta=1}^\alpha\ket{\beta}\bra{\beta}\nonumber \\&&-\alpha\ket{\alpha+1}\bra{\alpha+1}),
\end{eqnarray}
for $1\leq\alpha<\mathcal{N}$
\begin{eqnarray}
    \Lambda^i_{\mathcal{D}}&=&\sqrt{\frac{1}{\mathcal{D}}} \mathbb{I}.
\end{eqnarray}
With these operators, the local density matrix $\hat\rho_i$, as well as any operator $\hat O^i$ of local observables can be represented as
    \begin{eqnarray}
      \hat O^i&=&\sum_\mu c^i_\mu \Lambda^i_\mu, ~~~\text{with}\\
      c^i_\mu&=&{\rm Tr}[\Lambda^i_\mu\hat O^i],
    \end{eqnarray}
    and  $\mu=1,2,...\mathcal{D}$. This allows expressing both one-body and two-body Hamiltonians  in the form $ \hat H_i=\sum_\mu c^i_\mu\Lambda^i_\mu$, and $\hat H_{ij}=\sum_{\mu,\nu}c^{ij}_{\mu\nu}\Lambda^i_\mu\Lambda^j_\nu$.
The Heisenberg equations of motion for $\Lambda^i_\mu$ can be written as
    \begin{eqnarray}
      i\hbar\frac{d \Lambda^i_\mu}{dt}&=&[\Lambda^i_\mu,\hat H]\nonumber\\
                                             &=&\sum_\mu c^i_\nu[\Lambda^i_\mu,\Lambda^i_\nu]+\sum_{\sigma,j,\nu}c^{ij}_{\sigma,\nu}[\Lambda^i_\mu,\Lambda^i_\sigma]\Lambda^j_\nu.\label{eq:dLam}
    \end{eqnarray}
In the experiment, the initial state is a product state of single atom density matrices, $\hat\rho(t=0)=\prod\hat\rho^i(t=0)$. If we adopt a factorization $\langle \Lambda^i_\mu\Lambda^j_\nu...\Lambda^k_\sigma\rangle=\langle\Lambda^i_\mu\rangle\langle\Lambda^j_\nu\rangle...\langle\Lambda^k_\sigma\rangle$ for any non-equal $i,j,...k$ (i.\,e.\,each operator acts on a different atom) and arbitrary $\mu,\nu,\sigma$, Eq.\,S10 becomes  a closed set of  nonlinear equations for $\lambda^i_\mu=\langle\Lambda^i_\mu\rangle$. Within a mean-field treatment, the initial condition is fixed by $\lambda^i_\mu(t=0)={\rm Tr}[\Lambda^i_\mu\hat\rho(t=0)]$, which determines the ensuing dynamics from Eq.\,S10. This treatment neglects any correlations between atoms. In the  GDTWA method, the initial value of $ \lambda^i_\mu$  is instead sampled from  a probability distribution in  phase space, with statistical average  $\overline{\lambda^i_\mu(0)}={\rm Tr}[\Lambda^i_\mu\hat\rho(t=0)]$. Specifically, each $\Lambda^i_\mu$ can be decomposed via its eigenvalues and eigenvectors as $\Lambda^i_\mu=\sum_{a^i_\mu}a^i_\mu\ket{a^i_\mu}\bra{a^i_\mu}$. We take $a^i_\mu$ as the allowed values of $\Lambda^i_\mu$ in phase space, then for an initial state $\hat \rho^i(t=0)$, the probability distribution is  $p(a^i_\mu)={\rm Tr}[\hat \rho^i(t=0)\ket{a^i_\mu}\bra{a^i_\mu}]$.  From Eq.\,S10, each sampled initial configuration  for the $N$ atom array, $\{a_\mu\}=\{a^{i1}_{\mu_1},a^{i2}_{\mu_2},...a^{iN}_{\mu_N}\}$ leads to a  trajectory of $\Lambda^i_{\mu}$, which we denote as $\lambda^i_{\mu,\{a_\mu\}}(t)$. The quantum dynamics can be obtained by averaging over sufficient number of trajectories
\begin{eqnarray}
\lambda^i_\mu(t)&\approx&\overline{\lambda^i_\mu(t)}=\sum_{\{a_\mu\}}p(\{a_\mu\})\lambda^i_{\mu,\{a_\mu\}}(t).
\end{eqnarray}
This approach has been shown capable of capturing the buildup of quantum correlations~\cite{Schachenmayer2015b,Lepoutre2018}.

\section{Incorporating experimental conditions in numerical simulation}
In our experiment, the lattice filling fraction is not unity when the spin dynamics takes place. The reduced  filling fraction is due to two effects: the finite temperature and atom loss during the initial state preparation. To account for the effect of a finite temperature, we first obtain the density distribution before ramping  up the lattice from a Fermi-Dirac distribution $n^0({\bf r}_i)=\frac{1}{1+{\rm exp}(\beta(\epsilon({\bf r}_i)-\mu))}$, with parameters $\beta=1/k_BT$ and $\mu$ matching the inferred experiment temperature $T$, and the total atom number $N_0=2.4\times 10^4$. The function  $\epsilon({\bf r}_i)$ accounts for the weak external harmonic confinement. We compute the density distribution function after loading the atoms in the lattice, $n^F({\bf r}_i)$ by simulating the lattice ramp which is possible since to an excellent approximation we can treat the system as non-interacting. Indeed, we neglect the dipolar interaction in the loading given that their magnitude is much lower than the Fermi Energy of the gas. In the numerical simulation, we then sample the position of atoms ${{\bf r}_i}$ in the lattice according to a distribution $p({\bf r}_i)=n^F({\bf r}_i)/N_0$. In practice, to reduce computation cost we need to reduce the total atom number in our calculations and use a smaller lattice with fewer populated  lattice sites. In this case, we reduce the number of  lattice sites by a factor $\xi=(N_{sim}/N_{exp})^{1/3}$, where $N_{sim(exp)}$ are the number of atoms in  the simulation (experiment), while keeping the lattice spacings the same as in experiment, $(d_x,d_y,d_z)=(272,266,544)\,$nm. That is, for an initial lattice with $L_x$ sites along $x$ direction, in our simulations there are $\xi L_x$ sites while the separation between two adjacent lattice sites is still $d_x$. We then sample the initial  distribution of atoms in the lattice with $\tilde{p}(\tilde{\bf r}_i)=\xi^3p(\xi\tilde{\bf r}_i)$, which preserves the local density and is similar to sampling in a coarse-grained lattice. In our simulations, we chose $N_{sim}\gtrsim 350$ and  checked that the convergence in $N_{sim}$ has been reached.

\begin{figure}[t!]
 \centering
  \includegraphics[width=0.95\linewidth]{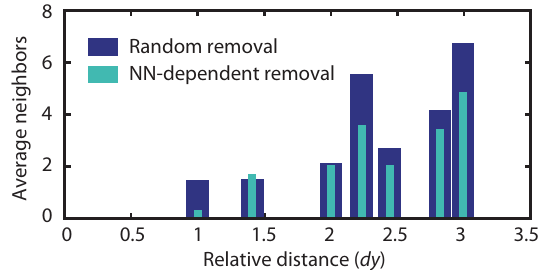}
\caption{Atom distribution Histogram showing the average number of atoms in distances normalized to the lattice direction along $y$ for random removal of atoms and for removal depending on the number of nearest neighbors (NN-dependent removal).}
\label{fig:S3}
\end{figure}

As discussed in Sec.\,S2, a fraction of atoms  
is lost during the ramp up and down of the magnetic field before initializing the spin dynamics over the sample. While a rigorous treatment on how these losses modify the distribution is not currently accessible with our current experimental setup, we try to account for it in the simulation by preferentially removing those atoms with a probablity $\propto p({\bf r}_i) N_{\rm nn}$, where $N_{\rm nn}$ is the number of nearest neighbors (separation $\le d_y$), until $N=\nu(0) N_0$ atoms are left. According to experiment estimates, the filling fractions before the initialization of the spin dynamics are $\nu(0) = 0.6 \sim 0.7$ (see Tab.\,S1~and~S2). Figure~\ref{fig:S3} shows the histogram of neighbors in the resulting atom distribution. Such distribution effectively reduces the nearest-neighbor interactions and is found to give a better agreement with experiment.  

Both the quadratic and linear shifts in the experiment are inhomogeneous across the lattice as discussed in Sec.\,S5, and we include them in our numerical simulation as site-dependent terms $\delta_{i} (\hat F_{i}^z)^2$ and $B_{ i}\hat F_{ i}^z$, with $\delta_{ i}=a |{\bf r}_i|^2$ and $B_{ i}=b(x_i+y_i+z_i)$. Based on experimental estimation, we have chosen the values of $a$ and $b$ such that $\delta_{i}=h\times 1.6\,$Hz ($h\times 0.7\,$Hz)  at 20 sites along $y$ away from the lattice center, and $B_{i}$ differs by $h\times 6\,$Hz ($h\times 1.8\,$Hz) between adjacent sites, for  Fig.\,\ref{fig:2}\, and\,\ref{fig:3} (Fig.\,\ref{fig:4})  in the simulation.

\section{Short-time population dynamics} Considering a fixed initial atomic distribution over the lattice, the population dynamics at early times can be derived  via a perturbative short time expansion 
\begin{eqnarray}
n_{m_F}(t)&\equiv& \langle\hat n_{m_F}(t)\rangle=\langle\hat n_{m_F}\rangle+i\langle[\hat H, \hat n_{m_F}]\rangle t/\hbar\nonumber\\
&&-\langle[\hat H,[\hat H,\hat n_{m_F}]]\rangle t^2/2\hbar^2\nonumber\\
&&-i\langle[\hat H,[\hat H,[\hat H, \hat n_{m_F}]]]\rangle t^3/3!\hbar^3  \nonumber\\
&&+\langle[\hat H,[\hat H,\allowbreak[\hat H,[\hat H,\hat n_{m_F}]]]\rangle t^4/4!\hbar^4\nonumber\\
&&+\mathcal{O}(t^5) 
\end{eqnarray} Here the average $\langle\cdot\rangle$ is over the initial state, which is assumed to be a pure state,  $\hat n_{m_F}=(\sum_{i} \mathcal{P}^{m_F}_i)/N$, %$\hat n_{m_F}=(\sum_{i=1}^N \mathcal{P}^{m_F}_i)/N$, 
where  $\mathcal{P}^{m_F}_i=|m_F\rangle_i{}_i\langle m_F| $ is the onsite projector for an atom at site $i$ in state $|m_F\rangle$ and $N$  denotes the total number of atoms.  Note that here the sums are always carried out over the populated lattice sites in the initial lattice configuration.  
We obtain
\begin{eqnarray}
  n_{m_F^0}(t)&=&n_{m_F^0}(0)\Big(1-n_{m_F^0}(0)\frac{V_{\rm eff}^2}{\hbar^2}t^2\nonumber\\
&&+\mathcal{O}(t^4)\Big),
\end{eqnarray}  
with
\begin{eqnarray} 
V_{\rm eff}^2&=&\frac{\gamma^2(m_F^0)}{8N}\sum_{{ i},{ j}\neq { i}} V_{{ i},{ j}}^2,
\end{eqnarray}
\begin{eqnarray}
  \gamma(m_F^0)&=&\sqrt{F(F+1)-m_F^0(m_F^0+1)} \nonumber \\
  && \times \sqrt{F(F+1)-m_F^0(m_F^0-1)},
\end{eqnarray}

where  $n_{m_F^0}$ denotes the population on the selected target state. To obtain Eq.\,S13, we have assumed that initially most of the population is in this target state, i.e. $n_{m_F^0}(0)\sim 1$. In the experiment, this assumption is always satisfied and therefore Eq.\,S15 is expected to reproduce well the short time dynamics.

The dependence of $\gamma(m_F^0)$ on the initial state $m_F^0$ is a consequence of the dependence of dipolar exchange processes on  the spin coherences, i.\,e.\,  $|\langle i:m_{F}^0+1,j:m_{F}^0-1| \hat F_i^+\hat F_j^-|i:m_{F}^0,j:m_{F }^0\rangle|$. Therefore the smaller the value  $|m_F^0|$  of the  initial populated  states, the  faster the early time  dynamics. Notably, up to  order $t^2$ the initial dynamics is independent of quadratic shifts and external magnetic field gradients. This is  because both of their corresponding Hamiltonians commute with the spin population operator $\hat n_{m_F}$. From this simple perturbative treatment  one learns that by preparing different initial states with different  $m_F^0$, the decay rates of the short time population dynamics provide information of $V_{\rm eff}$ and thus of the underlying dipolar couplings.
As discussed in Sec.\,S2 and S8, the lattice filling fraction is not unity and the initial atomic density distribution in the lattice may vary from shot to shot. To account for this effect, we perform a statistical average of Eq.\,S14 calculated for each lattice configuration generated with the procedure in Sec.\,S8 to obtain the theoretical values in Fig.\,\ref{fig:3}G and Fig.\,\ref{fig:4}B.   

\begin{figure}[t!]
 \centering
  \includegraphics[width=0.95\linewidth]{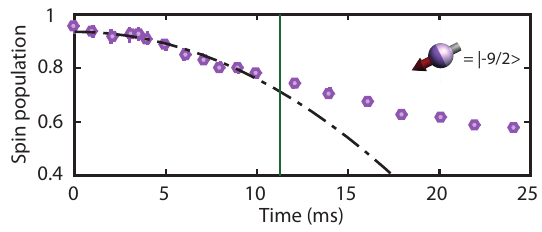}
    \caption{Fitting example to extract $V_{\rm{eff}}$. The dotted-dashed line exemplary shows the fit of Eq.\,S13 to the experimental data to extract $V_{\rm{eff}}$ for $|m_F^0 \rangle = \mnine$. The solid green line indicates the time $t_{\rm{fit}}$ up to which the fit is performed.}
\label{fig:S4}
\end{figure}

It is important here to compare the predictions obtained from a simple mean-field analysis. In contrast to Eq.\,S15, neglecting quantum correlations yields 
\begin{eqnarray}
  n_{m_F^0}^{\rm Mean-Field}(t)&=&n_{m_F^0}(0)\Big(1-n_{m_F^0}(0)[1-n_{m_F^0}(0)]\frac{V_{\rm eff}^2}{\hbar^2}t^2\nonumber\\ &&+\mathcal{O}(t^4)\Big ). \label{eq:shortT:mf}
\end{eqnarray} At the mean field level therefore if initially  the atoms are prepared such that $n_{m_F^0}(0)=1$, then there is no  population dynamics. This is in stark contrast to the quantum systems where dynamics is enabled by quantum fluctuations.
To extract $V_{\rm{eff}}$ from our experimental data and to compare it to the theoretical simulations we fit the initial dynamics with Eq.\,S15. We define the time scale for the fitting via $t_{\rm{fit}} < 0.5\frac{\hbar}{V_\text{eff}}$, which corresponds to the timescale on which each atom did on average half a spin flip. We note that on this timescale the time evolution starts already to deviate from the short time expansion (Eq.\,2), leading to a systematic downshift of the experimentally fitted $V_\text{eff}$; see Fig.\,\ref{fig:4}B. However, a minimum timescale has to be chosen to ensure that the fit is performed using a large enough number of datapoints. Figure~\ref{fig:S4} shows exemplary the fit to the experimental data for $|m_F^0 \rangle= \mnine$.

\end{document}